\documentclass[aps,twocolumn,prd,preprintnumbers, 10pt]{revtex4-1}
\pdfoutput=1
\usepackage{color}
\usepackage{enumitem}
\usepackage[dvipsnames]{xcolor}
\usepackage{hyperref}
\definecolor{amaranth}{rgb}{0.9, 0.17, 0.31}
\definecolor{forestgreen(web)}{rgb}{0.13, 0.55, 0.13}
\definecolor{blue(munsell)}{rgb}{0.0, 0.5, 0.69}
\definecolor{bblue}{rgb}{0.0, 0.58, 0.71}
\hypersetup{
pdfstartview={FitH}, 
pdftitle={}, 
colorlinks=true, 
linkcolor=blue(munsell), 
citecolor=blue, 
filecolor=magenta, 
urlcolor=blue 
}
\usepackage{pgfplots}
\usepackage{amssymb}
\usepackage{amsfonts}
\usepackage{graphicx}
\usepackage{epstopdf}
\usepackage{dcolumn}
\usepackage{amsmath}
\usepackage{latexsym,bm}
\usepackage{amsthm}
\usepackage{slashed}
\usepackage{float}
\usepackage{color}
\usepackage{url}
\usepackage{longtable}
\usepackage{rotating}
\usepackage[normalem]{ulem}
\usepackage{faktor}
\usepackage{tikz-cd}
\usepackage{tensor}
\usepackage{array}
\usepackage{tabularx}
\usepackage{makecell}
\usepackage{changepage}

\usepackage{tikz}
\usetikzlibrary{arrows.meta}
\usepackage{diagbox}

\usepackage[english]{babel} 
\usepackage[autostyle]{csquotes}
\usepackage{pifont}
\usepackage{xstring}
\usetikzlibrary{decorations.pathmorphing} 
\usetikzlibrary{decorations.markings} 
\usetikzlibrary{arrows} 
\usetikzlibrary{shapes} 
\usetikzlibrary{matrix} 
\usetikzlibrary{positioning} 
\usetikzlibrary{positioning}
\usetikzlibrary{calc}
\usetikzlibrary{decorations.pathreplacing,calligraphy}

\usetikzlibrary{shapes.multipart}

\tikzset{->-/.style={decoration={
  markings,
  mark=at position .5 with {\arrow{>}}},postaction={decorate}}}
  \usepackage{tikz,tikz-3dplot,tikz-cd,circuitikz}
\usetikzlibrary{patterns,decorations.markings, decorations.pathmorphing, arrows,calc,knots,hobby,
	decorations.pathreplacing,
	shapes.geometric,
	calc, arrows, arrows.meta, decorations.text}

\allowdisplaybreaks

\usepackage{tabularx}
\usepackage{siunitx,booktabs,array}

\newcommand{\be}{\begin{equation}}
\newcommand{\ee}{\end{equation}}
\newcommand{\ba}{\begin{aligned}}
\newcommand{\ea}{\end{aligned}}

\definecolor{DarkGreen}{rgb}{0.1, 0.7, 0.3}

\newcommand{\Rep}{\mathsf{Rep}}

\newcommand{\bit}{\begin{itemize}}
\newcommand{\eit}{\end{itemize}}
\newcommand{\ben}{\begin{enumerate}}
\newcommand{\een}{\end{enumerate}}

\renewcommand{\ni}{\noindent}

\newcommand{\wt}{\widetilde}

\newcommand{\half}{\frac{1}{2}}

\newcommand{\Z}{{\mathbb Z}}

\newcommand{\bC}{{\mathbb C}}

\newcommand{\C}{\mathsf{C}}

\newcommand{\expOP}[2]{
\langle #1|#2|#1\rangle}

\def\half{{\frac{1}{2}}}

\def\unit{{1\kern-.65ex {\rm l}}}
\def\1{{1\kern-.65ex {\rm l}}}

\newcommand{\beq}{\begin{equation}}
\newcommand{\eeq}{\end{equation}}

\usepackage{verbatim}
\usepackage{tikz}
\usetikzlibrary{arrows,snakes,shapes.arrows,decorations.markings}
     \tikzset{>=triangle 90}
     \tikzstyle{gr}=[draw,circle,green!50!black,fill=green!50!black,scale=.6]
     \tikzstyle{Bl}=[draw,circle,blue,scale=.7]
     \tikzstyle{R}=[draw,circle,fill=red,scale=.7]
     \tikzstyle{bl}=[draw,circle,fill=black,scale=.2]
     \tikzstyle{bbc}=[draw,circle,fill=black,scale=.75]
     \tikzstyle{bbcs}=[draw,circle,fill=black,scale=.5]
     \tikzstyle{rc}=[circle,fill=red,scale=.6]
     \tikzstyle{wc}=[draw,circle,scale=.75]
\usepackage{array} 
\usepackage{multirow} 
\usepackage{colortbl} 
\usepackage{stfloats}  
\usepackage{cellspace}
\usepackage{xcolor}   

\newcommand{\xdasharrow}[2][->]{
\tikz[baseline=-\the\dimexpr\fontdimen22\textfont2\relax]{
\node[anchor=south,font=\scriptsize, inner ysep=1.5pt,outer xsep=2.2pt](x){#2};
\draw[shorten <=3.4pt,shorten >=3.4pt,dashed,#1](x.south west)--(x.south east);
}
}

\def\^{\wedge}

\def\C{\mathbb{C}}

\def\Z{\mathbb{Z}}

\newcount\hour \newcount\minute
\hour=\time \divide \hour by 60
\minute=\time
\def\now{%
\ifnum \hour<13
  \ifnum \hour=0 \advance \hour by 12 \number\hour:\else \number\hour:\fi%
     \ifnum \minute<10 0\fi%
     \number\minute%
\ A.M.%
\else \advance \hour by -12 \number\hour:%
  \ifnum \minute<10 0\fi%
  \number\minute%
  \ P.M.%
\fi%
}

\usepackage{tikz}
\usetikzlibrary{arrows}
\usetikzlibrary{shapes.geometric,calc,arrows, positioning,shapes.misc,decorations.markings}
\tikzset{
  big arrow/.style={
    decoration={markings,mark=at position 1 with {\arrow[scale=2,#1]{>}}},
    postaction={decorate},
    shorten >=0.4pt},
  big arrow/.default=black}
\usetikzlibrary{external}
\usetikzlibrary{positioning}
\usetikzlibrary{calc}
\tikzset{gauge-node/.style={shape=circle, draw, minimum width=.6cm}}

\pgfdeclarelayer{edgelayer}
\pgfdeclarelayer{nodelayer}
\pgfsetlayers{edgelayer,nodelayer,main} 
\tikzstyle{none}=[inner sep=0pt] 

\tikzstyle{NodeCross}=[draw, shape=circle, cross out, inner sep=0pt, minimum size=6pt,line width=0.25mm]
\tikzstyle{Circle}=[draw, shape=circle, black, inner sep=0pt, minimum size=6pt]
\tikzstyle{rtriangle}=[fill=black, regular polygon, regular polygon sides=3, rotate=90, inner sep=0pt, minimum size=8pt]
\tikzstyle{ltriangle}=[fill=black, regular polygon, regular polygon sides=3, rotate=270, inner sep=0pt, minimum size=8pt]
\tikzstyle{rtriangleblue}=[fill={rgb,255: red,17; green,160; blue,255}, regular polygon, regular polygon sides=3, rotate=90, inner sep=0pt, minimum size=8pt]
\tikzstyle{ltriangleblue}=[fill={rgb,255: red,17; green,160; blue,255}, regular polygon, regular polygon sides=3, rotate=270, inner sep=0pt, minimum size=8pt]
\tikzstyle{rtrianglegreen}=[fill={rgb,255: red,69; green,255; blue,28}, regular polygon, regular polygon sides=3, rotate=90, inner sep=0pt, minimum size=8pt]
\tikzstyle{ltrianglegreen}=[fill={rgb,255: red,69; green,255; blue,28}, regular polygon, regular polygon sides=3, rotate=270, inner sep=0pt, minimum size=8pt]
\tikzstyle{Uprtriangle}=[fill=black, regular polygon, regular polygon sides=3, rotate=0, inner sep=0pt, minimum size=8pt]
\tikzstyle{Downltriangle}=[fill=black, regular polygon, regular polygon sides=3, rotate=180, inner sep=0pt, minimum size=8pt]
\tikzstyle{rtriangleAmber}=[fill={rgb,255: red, 191; green, 144; blue, 63}, regular polygon, regular polygon sides=3, rotate=90, inner sep=0pt, minimum size=8pt]
\tikzstyle{UprtriangleViolett}=[fill={rgb,255: red,255; green,0; blue,0}, regular polygon, regular polygon sides=3, rotate=0, inner sep=0pt, minimum size=8pt]

\tikzstyle{Downltriangle}=[fill=black, regular polygon, regular polygon sides=3, rotate=180, inner sep=0pt, minimum size=8pt]
\tikzstyle{UpRighttriangle}=[fill=black, regular polygon, regular polygon sides=3, rotate=45, inner sep=0pt, minimum size=8pt]
\tikzstyle{UpLefttriangle}=[fill=black, regular polygon, regular polygon sides=3, rotate=315, inner sep=0pt, minimum size=8pt]
\tikzstyle{DownRighttriangle}=[fill=black, regular polygon, regular polygon sides=3, rotate=135, inner sep=0pt, minimum size=8pt]
\tikzstyle{DownLighttriangle}=[fill=black, regular polygon, regular polygon sides=3, rotate=225, inner sep=0pt, minimum size=8pt]

\tikzstyle{Star}=[draw, shape=star, fill=black, star points=8, inner sep=0pt, minimum size=8pt]

\tikzstyle{DashedLine}=[-, densely dashed, line width=0.25mm]
\tikzstyle{DashedLineBrown}=[-, densely dashed, line width=0.25mm, draw={rgb,255: red,155; green,103; blue,51}]
\tikzstyle{DashedLineFall}=[-, densely dashed, line width=0.25mm, draw={rgb,255: red,195; green,0; blue,0}]
\tikzstyle{DashedLineViolett}=[-, densely dashed, line width=0.25mm, draw={rgb,255: red,139; green,41; blue,148}]
\tikzstyle{DottedLine}=[-, dotted, line width=0.25mm]
\tikzstyle{BlueLine}=[-, fill=none, draw={rgb,255: red,17; green,160; blue,255}, line width=0.25mm]
\tikzstyle{GreenLine}=[-, fill=none, draw={rgb,255: red,69; green,255; blue,28}, line width=0.25mm]
\tikzstyle{RedLine}=[-, draw={rgb,255: red,191; green,0; blue,0}, fill=none, line width=0.25mm]
\tikzstyle{DashedLineRed}=[-, densely dashed, fill=none, draw={rgb,255: red,191; green,0; blue,0}, line width=0.25mm]
\tikzstyle{ThickLine}=[-, line width=0.25mm]
\tikzstyle{ViolettLine}=[-, draw={rgb,255: red,132; green,60; blue,191}, fill=none, line width=0.25mm]
\tikzstyle{ViolettDashedLine}=[-, densely dashed, draw={rgb,255: red,132; green,60; blue,191}, fill=none, line width=0.25mm]
\tikzstyle{AmberLine}=[-, draw={rgb,255: red,191; green,144; blue,63}, fill=none, line width=0.25mm]
\tikzstyle{DashedRedThick}=[-, densely dashed, fill=none, draw={rgb,255: red,191; green,0; blue,0}, line width=0.40mm]
\tikzstyle{DashedBlueThick}=[-, densely dashed, fill=none, black, line width=0.40mm]

\makeatother

\begin{document}

\title{Illustrating the Categorical Landau Paradigm in Lattice Models}

\author{Lakshya Bhardwaj$^1$}
\author{Lea E.\ Bottini$^1$}
\author{Sakura Sch\"afer-Nameki$^1$}
\author{Apoorv Tiwari$^2$}

\affiliation{${}^1$Mathematical Institute, University
of Oxford, Woodstock Road, Oxford, OX2 6GG, United Kingdom}
\affiliation{${}^2$Niels Bohr International Academy, Niels Bohr Institute, University of Copenhagen, Denmark}


\begin{abstract} 
\noindent 
Recent years have seen the concept of global symmetry extended to non-invertible (or categorical) symmetries, for which composition of symmetry generators is not necessarily invertible. 
Such non-invertible symmetries lead to a generalization of the standard Landau paradigm.  In this work we substantiate this framework by providing a (1+1)d lattice model, whose gapped phases and phase transitions can only be explained by symmetry breaking of non-invertible symmetries. 
\end{abstract}


\maketitle


\ni\textbf{Introduction.}
Global symmetries play a crucial role in our understanding of physics as they provide powerful constraints on the dynamics of a theory; for instance, they restrict the possible phases a theory can flow to in the infra-red (IR). Not only can the IR gapped phases of a theory with global symmetry  be organized in terms of patterns of spontaneous symmetry breaking (SSB), the phase transitions between such phases are also constrained by the symmetry. When the symmetry is a group $G$, this is the content of the  Landau paradigm. 

Starting with \cite{Gaiotto:2014kfa}, the concept of symmetries has been vastly generalized, including the extension to so-called  {\it categorical} or {\it non-invertible} symmetries \footnote{See \cite{Schafer-Nameki:2023jdn, Shao:2023gho} for recent reviews on the topic.}. As the name implies, invertibility of the symmetry is relaxed, and  schematically the composition takes the form
$a \otimes b = \sum_{c} n_{ab}^c c$,
where the coefficients are non-negative integers, and $c$ is summed over all symmetry generators.

In light of this, one is naturally led to explore the constraints of such symmetries on IR physics.  In recent works \cite{Bhardwaj:2023fca,Bhardwaj:2023idu,Bhardwaj:2023bbf,Bhardwaj:2024qrf} \footnote{For related works using similar SymTFT approaches for invertible symmetries to determine gapped phases, see \cite{Vanhove:2018wlb, Chatterjee:2022jll, Chatterjee:2022tyg, Wen:2023otf, Huang:2023pyk, Moradi:2022lqp, Chen:2022wvy}. For gapped phases with non-invertible symmetries in lattice models see \cite{Fechisin:2023dkj,Seifnashri:2024dsd}.},
a proposal for studying gapped and gapless phases protected by categorical symmetries was put forward in Quantum Field Theory (QFT) using the so-called Symmetry Topological Field theory (SymTFT) \cite{Ji:2019jhk, Gaiotto:2020iye, Apruzzi:2021nmk, Freed:2022qnc}.  
This extends the standard Landau theory  to a ``categorical Landau paradigm" \cite{Bhardwaj:2023fca}, leading to new phases and phase transitions.  

The goal of this work is to show, by means of  a  concrete lattice model, how this extension to non-invertible symmetries provides a crucial theoretical tool to understand beyond Landau phases. 
Our lattice model is realized on a tensor product Hilbert space, acted upon by generalized Ising Hamiltonians. These models exhibit four gapped phases, with a commuting projector Hamiltonian within each of them. The ground states cannot be explained as standard SSB phases, but require a non-invertible symmetry, in this case Rep$(S_3)$, which is generated by irreducible representations of the permutation group $S_3$. Moreover, by tuning the parameters in the generalized Ising Hamiltonians, we also realize second order phase transitions between such gapped phases. The order parameters for the phase transitions are mixtures of local and string-like order parameters, which is a hallmark of non-invertible symmetries \cite{Bhardwaj:2023ayw,Bhardwaj:2023idu}. 

Interestingly, we find a phase transition involving two degenerate gapless states, whose degeneracy also cannot be explained by standard SSB, but only by invoking the non-invertible $\Rep(S_3)$ symmetry. The gapless systems providing this phase transition form what is known as an intrinsically gapless SSB (igSSB) phase \cite{Bhardwaj:2024qrf} that exhibits \textit{symmetry protected criticality}: any symmetric deformation of the system either preserves the two degenerate gapless states, or gaps the system in such a way that the number of degenerate ground states necessarily increase, thus increasing the amount of order.

This lattice model provides a concrete ultra-violet (UV) realization of the gapped and gapless $\Rep(S_3)$ phases found using continuum methods in \cite{Bhardwaj:2023idu,Bhardwaj:2023bbf,Bhardwaj:2024qrf}. We remark that lattice models whose phases can be characterized in terms of a $\Rep(S_3)$ symmetry have been discussed also in \cite{Eck:2023gic,eck2023critical}, with the key difference that such models are realized on constrained Hilbert spaces, and not on a tensor product Hilbert space like the one presented here.




\begin{figure}[H]\centering
\begin{tikzpicture}
\draw [yellow,  fill=yellow, opacity = 0.1]   
(-1.5, -2) -- (7, -2) -- (7, -5.5) -- (-1.5,-5.5) --(-1.5,-2) ;
\draw [cyan,  fill=cyan, opacity = 0.1]   
(7, 1.5) -- (7, -5.5) -- (2.5,-5.5) --(2.5,1.5) -- (7,1.5);
\node[fill= white, draw,rectangle,thick,align=center](Trivial) at 
(0.5,0) 
{{I: Trivial} \\  
\begin{tikzpicture}
\node at (2,0) {${{\rm GS}}$};
\node[purple] at (3.5,0) {$\Rep(S_3)$};
\draw [purple, -stealth](2.4,-0.2) .. controls (2.9,-0.5) and (2.9,0.4) .. (2.4,0.1);
\end{tikzpicture} };
\node[fill= white,draw,rectangle,thick,align=center](RepS3Z2) at 
(5,0) 
{{III: $\Rep (S_3)/\Z_2$ SSB} \\ 
\centering
\begin{tikzpicture}
\begin{scope}[shift={(-1,0)}]
\node at (0,0) {${{\rm GS}_0}$};
\node at (1,0) {${{\rm GS}_1}$};
\node at (2,0) {${{\rm GS}_2}$};
\draw [red, stealth-stealth](0.2,-0.3) .. controls (0.4,-0.6) and (0.7,-0.6)    .. (0.9,-0.3);
\begin{scope}[shift={(0.9,0)}]
\draw [red, stealth-stealth](0.2,-0.3) .. controls (0.4,-0.6) and (0.7,-0.6)    .. (0.9,-0.3);
\end{scope}
\draw [red, stealth-stealth](-0.1,-0.3) .. controls (0.4,-1) and (1.5,-1) .. (2,-0.3);
\draw [blue, -stealth](0,0.3) .. controls (-0.4,0.9) and (0.4,0.9)   .. (0.1,0.3); 
\draw [blue, -stealth](1,0.3) .. controls (0.6,0.9) and (1.4,0.9)   .. (1.1,0.3); 
\draw [blue, -stealth](2,0.3) .. controls (1.6,0.9) and (2.4,0.9)   .. (2.1,0.3); 
\end{scope}
\end{tikzpicture}
};
\node[fill= white, draw,rectangle,thick,align=center](Z2SSB) at (0.5,-4) {{II: $\Z_2$ SSB} \\  
\begin{tikzpicture}[baseline]
\node at (0.7,0) {${{\rm GS}^+}$};
\node at (2.4,0) {${{\rm GS}^-}$};
\draw [red, stealth-stealth](0.7,-0.3) .. controls (0.9,-0.7) and (2.2,-0.7) .. (2.4,-0.3);
\begin{scope}[shift={(3,0)}]
\draw [rotate =180, blue, stealth-stealth](0.7,-0.3) .. controls (0.9,-0.7) and (2.2,-0.7) .. (2.4,-0.3);
\end{scope}
\begin{scope}[shift={(0.3, 0)}]
\draw [red, -stealth](2.4,-0.2) .. controls (2.9,-0.5) and (2.9,0.4) .. (2.4,0.1);
\end{scope}
\begin{scope}[shift={(2.7, 0)}]
\draw [rotate=180, red, -stealth](2.4,-0.2) .. controls (2.9,-0.5) and (2.9,0.4) .. (2.4,0.1);
\end{scope}
\end{tikzpicture}
};
\node[fill= white,draw,rectangle,thick,align=center](RepS3) at (5,-4) {{IV: $\Rep(S_3)$ SSB} \\ 
\begin{tikzpicture}[baseline]
\node at (0,0) {${{\rm GS}_2}$};
\node at (1,0) {${{\rm GS}_1}$};
\node at (2,0) {${{\rm GS}_0}$};
\draw [red, stealth-stealth](0,-0.3) .. controls (0.25,-0.7) and (0.75,-0.7)    .. (1,-0.3);
\draw [red, stealth-stealth](-0.1,-0.3) .. controls (0.5,-1) and (1.5,-1) .. (2,-0.3);
\draw [blue, -stealth](0,0.3) .. controls (-0.4,0.9) and (0.4,0.9)   .. (0.1,0.3); 
\draw [blue, stealth-stealth](1,0.3) .. controls (1.25,0.9) and (1.75,0.9)   .. (2,0.3); 
\begin{scope}[shift={(2, 0)}]
\draw [rotate=180, red, -stealth](2.4,-0.2) .. controls (2.9,-0.5) and (2.9,0.4) .. (2.4,0.1);
\end{scope}
\end{tikzpicture}
};
\draw[thick,stealth-stealth] (Trivial) to (RepS3Z2);
\draw[thick,stealth-stealth] (RepS3) to (RepS3Z2);
\draw[thick,stealth-stealth] (Trivial) to (Z2SSB);
\node at (2.6,-0.5) {Potts};
\draw [rotate=-90, purple, -stealth](0.7,2.4) .. controls (1.2,2.1) and (1.2,3.1) .. (0.7,2.8);
 \node[right] at (0.5,-2) {Ising};
\draw [rotate=0, purple, -stealth](1.4,-2.2) .. controls (1.9,-2.5) and (1.9,-1.6) .. (1.4,-1.9);
  \node[left] at (5,-2) {Ising$\oplus$Ising};
\draw [rotate=90, blue, -stealth](-1.8,-3.7) .. controls (-1.3,-4) and (-1.3,-3.1) .. (-1.8,-3.4);
\begin{scope}[shift={(1,-0.1)}]
\draw [rotate=90, blue, -stealth](-1.7,-3.7) .. controls (-1.2,-4) and (-1.2,-3.1) .. (-1.7,-3.4);
\end{scope}
\draw [rotate=180, red, -stealth](-3.1,1.9) .. controls (-2.6,1.5) and (-2.6,2.4) .. (-3.1,2.1);
\begin{scope}[shift={(2.1,-5.3)}]
\draw [rotate=-90, red, stealth-stealth](-3.1,1.4) .. controls (-2.7,1.6) and (-2.7,2.2) .. (-3.1,2.4);
\end{scope}
\end{tikzpicture}
\caption{The Hamiltonian (\ref{eq:Ham_3site}) has four gapped phases.  These are explained by the $\Rep(S_3)$ non-invertible symmetry breaking, whose action on the gapped ground states GS and gapless states is shown in blue (for symmetry operator $U$) and red (for symmetry operator $E$), with purple showing the full $\Rep(S_3)$ action. The phase transitions are indicated by black arrows. Non-zero vevs of order parameters are shaded yellow (for $O_q$) and blue (for $O_p$), with their intersection region shaded green.}
\label{fig:NotHasse}
\end{figure}
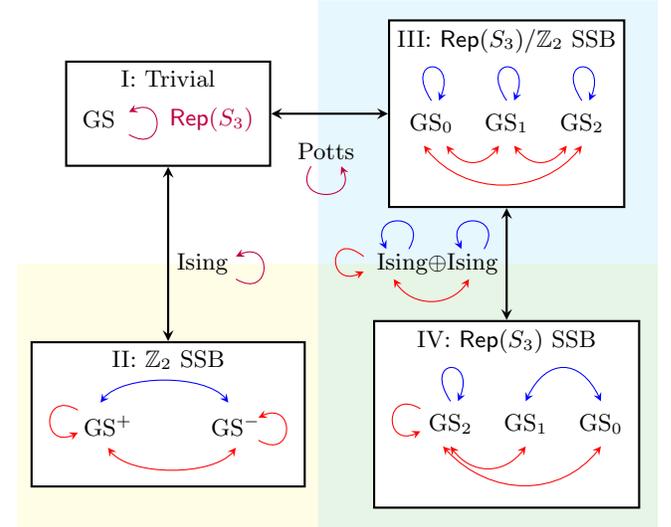

\ni\textbf{Model.}
We label the sites by integers and half-integers. On each integer site $j \in \Z$ we have a qutrit realizing $\bC^3$, and on half-integer sites $j+\half$, a qubit realizing $\bC^2$. 
 A basis state of the full system is labeled as
\be
|\vec p,\vec q\,\rangle=|\cdots,p_j,q_{j+\half},p_{j+1},q_{j+\frac32},\cdots\rangle
\label{eq:basis}
\ee
with $p_j\in\{0,1,2\}$ and $q_{j+\half}\in\{0,1\}$.
Consider a space of generalized Ising Hamiltonians, on a length $L$ lattice, comprising of terms with 3-site interactions of the following form:
\vspace{-10pt}
\be
\vspace{-10pt}
H=-\sum_{j=1}^{L}\sum_{I=0}^5 \left[ \lambda_{I} P^{(I)}_{j-\frac{1}{2}}+t_{I}X^{(I)}_{j}\right]\,,
\label{eq:Ham_3site}
\ee
where the operators act on $(\C^3 \otimes \C^2)^L$ as 
\be
\begin{tikzpicture}
   \newcommand\Square[1]{+(-#1,-#1) rectangle +(#1,#1)};
\draw (0,0) -- (8,0);
 \draw [black,fill= black] (1, 0) ellipse (0.1 and 0.1);
     \draw[black, fill= black] (2,0) \Square{0.1} ;    
  \draw [black,fill= black] (3, 0) ellipse (0.1 and 0.1);
     \draw[black, fill= black] (4,0) \Square{0.1} ;   
   \draw [black,fill= black] (5, 0) ellipse (0.1 and 0.1);
        \draw[black, fill= black] (6,0) \Square{0.1} ;   
    \draw [black,fill= black] (7, 0) ellipse (0.1 and 0.1);
 \node[above] at (2,0.25) {$P$};
    \draw[black, dashed] (0.6,-0.25) rectangle ++(2.7,0.5);
\begin{scope}[shift={(3,0)}]
 \node[above] at (2,0.25) {$X$};
    \draw[black, dashed] (0.6,-0.25) rectangle ++(2.7,0.5);
\end{scope}
\end{tikzpicture}
\ee
Restricted to the qubits (squares), the $P$ and $X$ operators implement a disordering and ordering (in the $x$-basis) respectively, while restricted to the qutrits (dots), these implement an ordering and disordering respectively: 
\be
\begin{split}
\hspace{-30pt}
P_{j+\half}^{(2p+q)}&=\frac{1}{6}\left[ 1+(-1)^q\sigma^{z}_{j+\half}\right]\left[\sum_{n=0}^{2}\omega^{-pn}Z^n_{j}Z_{j+1}^{(2q-1)n}\right] \\
X^{(2p+q)}_{j}&=\left(X_j\right)^p \left(\sigma^{x}_{j-\half}\Gamma_j\sigma^x_{j+\half}\right)^q\,, 
\end{split}
\ee
for $p\in\{0,1,2\}$, $q\in\{0,1\}$ and $\omega=\exp(2\pi i/3)$. The local operators $\sigma^{\mu}_{j+1/2}$ are the usual Pauli operators, whereas the operators acting on the qutrit degrees of freedom are
$Z={\rm diag}\left(1, \omega, \omega^2\right)$ and 
\be
X=
\begin{pmatrix}
0 & 0 & 1 \\
1 & 0 & 0 \\
0 & 1 & 0 \\
\end{pmatrix},
\qquad 
\Gamma=
\begin{pmatrix}
1 & 0 & 0 \\
0 & 0 & 1 \\
0 & 1 & 0 \\
\end{pmatrix}\,.
\ee
These models exhibit four different gapped phases, which each have a commuting projector Hamiltonian realizing ground states that are (unitarily equivalent to) tensor product states. These can be used to extract the universal properties of the phase. 
We now describe the phases:

\vspace{1mm}

\ni\textbf{Phase I with One Ground State.} A representative commuting projector Hamiltonian within this gapped phase is provided by setting \footnote{We have chosen to normalize the terms such that each term is separately a projection operator, i.e., it has eigenvalues 0 and 1.} $\lambda_I=3t_I= 1$ for $I=0,2,4$ and $\lambda_I=t_I= 0$ otherwise 
in  \eqref{eq:Ham_3site}. Doing so, the Hamiltonian simplifies to 
\begin{equation}
    H_1= -\sum_{j}\left[\frac{1+\sigma^z_{j+\half}}{2}+ \frac{1+X_j+X^2_j}{3}\right]\,.
\end{equation}
The first term projects onto $\sigma^z=+1$ state for each qubit and the second term projects onto the $X=+1$ state for each qutrit. Thus, the ground state is a product state
\begin{equation}
    |{\rm GS}_1\rangle=\bigotimes_{j}\Big|X_{j}=1\,,\sigma^{z}_{j+\half}=1\Big\rangle=\frac{1}{3^{L/2}}\sum_{\vec{p}}\big|\vec{p},\vec{0}\big\rangle\,.
    \label{eq:GS_1}
\end{equation}

\ni\textbf{Phase II with Two Ground States.} A commuting projector Hamiltonian is given by setting $\lambda_I=6t_I=1$ for all $I$ in \eqref{eq:Ham_3site}. We note that 
$\sum_{I}P^{(I)}=1$, therefore the Hamiltonian simplifies to 
\begin{equation}
    H_{2}=-\sum_{j}\frac{1}{6}\left(1+X_{j}+X_{j}^2\right)\left(1+\sigma^{x}_{j-\half}\Gamma_j\sigma^{x}_{j+\half}\right)\,.
\end{equation}
The terms within the two parenthesis commute with one another therefore we may first project onto the qutrit states with $X_{j}=1$, energetically satisfying the operator in the first parenthesis. Since, $\Gamma_j=1$ on the qutrit state with $X_j=1$, we effectively need to satisfy $\sigma^x_{j-1/2}\sigma^x_{j+1/2}=1$ for each $j$. There are two ground states
\begin{equation}
\ba
|{\rm GS}_2,\pm\rangle
=\frac1{6^{L/2}}\sum_{\vec p,\vec q}(\pm1)^{\sum_j q_{j+\frac{1}{2}}}\big|\vec p,\vec q\big\rangle\,. 
    \ea
    \label{eq:GS_2}
\end{equation}

\ni\textbf{Phase III with Three Ground States.} For this phase we set  $\lambda_I=t_{I}=\delta_{I,0}$, resulting in
\begin{equation}
    H_3=-\sum_j\frac{1}{6}\left[ 1+\sigma^{z}_{j+\half}\right]\left[\sum_{n=0}^{2} Z^n_{j}Z_{j+1}^{-n}\right]\,,
\end{equation}
which simultaneously projects onto the $\sigma^{z}_{j+1/2}=1$ qubit states and $Z_jZ^{-1}_{j+1}=1$ quitrit states for all $j$. We thus find three ground states labeled by $n\in\{0,1,2\}$
\begin{equation}
    |{\rm GS}_3,n\rangle=\bigotimes_{j}\Big|Z_{j}=e^{\frac{2\pi i n}{3}}\,,\sigma^{z}_{j+\half}= 1\Big\rangle=\big|\vec{n}\,,\vec{0}\big\rangle\,.
    \label{eq:GS_3}
\end{equation}

\ni\textbf{Phase IV with Three Ground States.} Finally, consider $\lambda_I=2t_{I}=\delta_{I,0}+\delta_{I,1}$, resulting in
\begin{equation}
\begin{split}
    H_4&=-\half\sum_j \left(1+\sigma^{x}_{j-\half}\Gamma_j\sigma^x_{j+\half}\right) \\ 
    & \ -\frac{1}{6}\sum_{j}\sum_{\alpha=\pm 1}\left[ 1+\alpha\sigma^{z}_{j+\half}\right]\left[\sum_{n=0}^{2} Z^n_{j}Z_{j+1}^{-\alpha n}\right]    \,.
\end{split}
\end{equation}
This has three ground states  (see appendix \ref{H4_ground_states} for details)
%
\begin{equation}
\begin{split}
   |{\rm GS}_4,0\rangle&= \frac{1}{2^{L/2}}\sum_{\vec{q}}|\vec{0}\,,\vec{q}\rangle\,,  \\
   |{\rm GS}_4,1\rangle&= \frac{1}{2^{L/2}}\sum_{\vec{q}}(-1)^{\sum_jq_{j+\half}}|\vec{0}\,,\vec{q}\rangle\,,\\  
    |{\rm GS}_4,2\rangle&= \frac{1}{2^{L/2}}{\sum_{\vec{p}\,,\vec{q}}}'
    |\vec{p}\,,\vec{q}\rangle \,,   
    \label{eq:GS_4}
\end{split}
\end{equation}
where $\sum'$ sums over $p_j\neq 0$ and $p_{j}+p_{j-1} \ {\rm mod} \ 2=q_{j-\half}$. 

\smallskip \ni\textbf{Analysis of Gapped Phases.} 
Typically, degenerate ground states in a gapped phase can be explained in terms of spontaneous breaking of a symmetry. 
There is an obvious $\Z_2$ symmetry of \eqref{eq:Ham_3site} that measures the total spin parity of all the qubits, generated by the unitary
\be\label{U}
U=\prod_{j}\sigma^z_{j+\half}\,.
\ee
The gapped phases I and II can be explained in terms of spontaneous breaking of this $\Z_2$ symmetry: 
 $|{\rm GS}_1\rangle$ preserves it, whereas the states $|{\rm GS}_2,\pm\rangle$ are degenerate and $\Z_2$ exchanges them
\footnote{Strictly speaking, these two ground states are degenerate only in the infinite size/thermodynamic limit as at finite volume, one may add $\Z_2$-symmetric terms to the Hamiltonian to create an energy gap between symmetric and anti-symmetric combinations of the ground states.}.

What is the explanation for the three-fold degenerate ground states in gapped phases III and IV? Although there is a $\Z_3$ symmetry of the commuting projector Hamiltonian $H_3$ generated by the unitary operator $\prod_{j}Z_j$, other Hamiltonians near $H_3$ in the parameter space of models \eqref{eq:Ham_3site} explicitly break this $\Z_3$ symmetry, without lifting the three-fold degeneracy of the ground state subspace, and thus is not the explanation.
We will show that this space of Hamiltonians \eqref{eq:Ham_3site} exhibits a $\Rep(S_3)$ non-invertible symmetry,  whose spontaneous breaking explains the gapped phases III and IV.

\smallskip\ni\textbf{$\Rep(S_3)$ Symmetry.}
$\Rep(S_3)$ has symmetry generators given by the irreducible representations of the permutation group $S_3$. There are two non-trivial such generators: one implemented by a unitary operator $U$ and the other implemented by a non-unitary operator $E$, with multiplication rules (which is just the tensor product of representations)
\be\label{fusion}
U^2=1\,,\quad U\times E=E\times U=E\,,\quad E^2 = 1+U+E\,.
\ee
Note that $U$ generates a $\Z_2$ subsymmetry of $\Rep(S_3)$, and $E$ is what is referred to as a non-invertible symmetry, as its inverse does not exist.  



In order for a system to realize $\Rep(S_3)$ symmetry, both these operators $U$ and $E$ have to commute with the Hamiltonian. Within our model, the $U$ operator is realized as in \eqref{U}, while the $E$ symmetry generator is
\be
\begin{split}
\hspace{-5pt}    E&= \frac{1}{2}\Big(1+\prod_j\sigma^z_{j+\half}\Big)\left(T_1+T_2\right)\\
\hspace{-5pt}    T_s&= \frac{1}{2}\prod_{j=1}^L 
\sum_{n=1,2}\Big[\Big(1\hspace{-2pt}+(-1)^{n+1}\hspace{-2pt}\prod_{i=0}^{j-1}\sigma^z_{i+\half}\Big)X^{ns}_{j}\Big] \,.
\end{split}
\label{RepS3_operators}
\ee
%
The reader can easily check that these operators \eqref{RepS3_operators} satisfy the multiplication rules \eqref{fusion}. Appendix \ref{commute} shows that $E$ commutes with the Hamiltonian \eqref{eq:Ham_3site}.

\smallskip \ni \textbf{$\Rep(S_3)$ Action on Gapped Phases.}
We now show that the ground states of the four gapped phases form irreducible representations \footnote{More precisely, the action of $\Rep(S_3)$ is irreducible on the ground states exhibiting cluster decomposition in the infinite volume limit (such ground states are also known as vacua). That is, one can generate all vacua starting from any one vacuum and acting on it by $\Rep(S_3)$ generators. All the ground states of gapped phases displayed in this paper are actually vacuum states.} 
of the $\Rep(S_3)$ symmetry (see Fig.~\ref{fig:NotHasse}), implying that the phases, including III and IV, can all be explained by spontaneous breaking patterns of $\Rep(S_3)$ symmetry. All the four possible symmetry breaking patterns for $\Rep(S_3)$ discussed from the point of view of SymTFT in \cite{Bhardwaj:2023idu} are realized in our model.

The ground state of phase I is invariant under the action of $\Rep(S_3)$ (up to scalars), and is thus the trivial phase for the $\Rep(S_3)$ symmetry. In contrast, the two ground states of phase II are exchanged by $U$ and the action of $E$ is
\be
E|{\rm GS}_2,{\pm}\rangle=|{\rm GS}_2,+\rangle+|{\rm GS}_2,-\rangle
\ee
Next, the ground states of phase III are invariant under $U$, but transform into each other by the $E$ action as 
\be
E|{\rm GS}_3,n\rangle=\sum_{m=1,2}\big|{\rm GS}_3,n+m~(\text{mod 3})\big\rangle
\label{eq:E_action Phase 3}
\ee
Finally, the action of $U$ exchanges ground states $|{\rm GS}_4,0\rangle$ and $|{\rm GS}_4,1\rangle$ of phase IV, while leaving $|{\rm GS}_4,2\rangle$ invariant, and the action of $E$ is as follows
\begin{equation}
\begin{split}
E|{\rm GS}_4,0\rangle&=E|{\rm GS}_4,1\rangle= |{\rm GS}_4,2\rangle \\
E|{\rm GS}_4,2\rangle&=|{\rm GS}_4,0\rangle+ |{\rm GS}_4,1\rangle+ |{\rm GS}_4,2\rangle\,.
\label{eq:E action on phase 4}
\end{split}
\end{equation}
Thus, the ground state degeneracy of both phases III and IV can be explained in terms of spontaneous breaking of the non-invertible symmetry $E$. The two phases are additionally distinguished by the fact that phase IV has ground states that also spontaneously break $\Z_2$ subsymmetry $U$, but all ground states of phase III preserve $U$. Phases III and IV were referred to as $\Rep(S_3)/\Z_2$ SSB and $\Rep(S_3)$ SSB phases respectively in \cite{Bhardwaj:2023idu}, and we will follow this nomenclature henceforth.

\smallskip\ni\textbf{Order Parameters.} The ground states of the four gapped phases can be distinguished by expectation values of the following two local order parameters
\begin{equation}
O_{q,j+\half}=\sigma^{x}_{j+\half}\,, \quad O_{p,j}=Z_{j}\,.
\end{equation}
We can easily compute their expectation values in the various ground states, which are
\be
\ba
\expOP{{\rm GS}_1}{O_{q}} &=0 \\ 
\expOP{{\rm GS}_2,\pm}{O_{q}}&=\pm 1\\
\expOP{{\rm GS}_3, n}{O_{q}}&=0 \\
\expOP{{\rm GS}_4, 0}{O_{q}}&=1\\
\expOP{{\rm GS}_4, 1}{O_{q}}&=-1\\
\expOP{{\rm GS}_4, 2}{O_{q}}&=0
\ea\quad
\ba
\expOP{{\rm GS}_1}{O_{p}}&=0 \\
\expOP{{\rm GS}_2,\pm}{O_{p}}&=0\\
\expOP{{\rm GS}_3, n}{O_{p}}&=e^{\frac{2\pi i n}{3}} \\
\expOP{{\rm GS}_4, 0}{O_{p}}&=1 \\
\expOP{{\rm GS}_4, 1}{O_{p}}&=1 \\
\expOP{{\rm GS}_4, 2}{O_{p}}&=-1/2\,.
\ea
\ee
The condensation of $O_q$, which is charged under the $\Z_2$ subsymmetry $UO_qU^{-1}=-O_q$,  characterizes the spontaneous breaking of $U$. 
On the other hand, in phases III and IV, $O_p$ is charged under $E$ and its condensation characterizes spontaneous breaking of $E$.
In the language of \cite{Bhardwaj:2023idu}, the operator $O_p$ lies in a multiplet carrying generalized charge \cite{Bhardwaj:2023ayw,Bhardwaj:2023wzd} $\mathbf{Q}_{[a],1}$ of the $\Rep(S_3)$ symmetry, discussed in more detail in the accompanying paper \cite{Bhardwaj:2024kvy}.

\smallskip\ni\textbf{Phase Transitions.}
To study phase transitions, we consider the simplest model of a one-parameter interpolation between two commuting projector Hamiltonians: e.g. for the transition between Phases $I$ and $J$ this is 
\vspace{-6pt}
\begin{equation}
    H_{I,J}(\lambda)=\lambda H_I + (1-\lambda)H_J\,.
\vspace{-7pt}
\end{equation}
One already encounters some interesting transitions in this simplistic space of models. For $H_{1,2}$, the low energy physics is within the $X_j=1$ subspace.  Noting that $\Gamma_j$ acts identically within this space, one obtains
\begin{equation}
    H_{1,2}(\lambda)\approx -\half \sum_j \left[\lambda \sigma^z_{j+\half}+(1-\lambda)\sigma^x_{j-\half}\sigma^x_{j+\half}\right]\,,
\end{equation}
where $\approx$ denotes that the Hamiltonian on the right hand side only describes the low energy physics of $H_{1,2}(\lambda)$. The critical Ising model describing the transition between gapped phases I and II is at $\lambda=1/2$, while the $\lambda>1/2$ and $\lambda<1/2$ regions corresponds to gapped phases I and II where the $U$ symmetry is preserved and spontaneously broken, respectively. The operator $O_q$ becomes the spin operator of the Ising model, which is the well-known order parameter for this transition.

Similarly, for $H_{1,3}$, the low-energy physics lies in the $\sigma^z=1$ subspace, in which we find the three-state Potts model spin chain Hamiltonian
\begin{equation}
H_{1,3}(\lambda)\approx -\frac{1}{3}\sum_{j}\sum_{n=0}^2\left[\lambda X_j^n+ (1-\lambda)Z_j^nZ_{j+1}^{-n}\right]\,.     
\end{equation}
This model has an emergent $\Z_3$ symmetry generated by $\eta=\prod_jX_j$, which is explained by the fact that at low energies, i.e., in the $\sigma^z=1$ subspace, the $E$ symmetry operator in \eqref{RepS3_operators} decomposes into $\eta+\eta^2$. The corresponding $\Z_3$ breaking transition occurs at $\lambda=1/2$. The operator $O_p$ becomes the spin operator of the Potts model, which is the standard order parameter for this transition.

Lastly, we discuss the transition between the two phases with three ground states modeled by $H_{3,4}$.
Note that the Hamiltonian $H_{3,4}$ block decomposes into two state spaces $V_1$ and $V_2$. $V_1$ is spanned by a basis $|\vec{0},\vec{q}\rangle$, while $V_2$ is spanned by states $|\vec{p},\vec{q}\rangle$ such that $p_j\neq 0$ and $q_{j+1/2}=p_{j+1}+p_j \ \text{mod 2}$. 
In the $V_1$ subspace, $Z_j=\Gamma_j=1$ for all $j$, therefore
\begin{equation}
\hspace{-5 pt}    H_{3,4}(\lambda)\Big|_{V_1}=-\frac{1}{2}\sum_j\left[ \lambda \sigma^z_{j+\half}+(1-\lambda)\sigma^x_{j-\half}\sigma^x_{j+\half}\right]\,.
\vspace{-5pt}
\label{eq:V1_Ham}
\end{equation}
For $V_2$, we define effective qubits $\widetilde{\sigma}^{\mu}_j$ such that the states $p_j=1,2$ are $\widetilde{\sigma}^{z}_j$ eigenstates with eigenvalues $+1$ and $-1$ respectively. In terms of these 
\begin{equation}
    \hspace{-5 pt}    H_{3,4}(\lambda)\Big|_{V_2}=-\frac{1}{2}\sum_j\left[ \lambda \widetilde{\sigma}^z_{j}\widetilde{\sigma}^z_{j+1}+(1-\lambda)\widetilde{\sigma}^x_{j}\right]\,.
\vspace{-5pt}
\label{eq:V2_Ham}
\end{equation}
Note that the $U$ symmetry acts trivially within $V_2$ and as the $\Z_2$ symmetry measuring spin parity within $V_1$. 
The action of the $E$ symmetry is more interesting as it maps between the dynamically disconnected state spaces $V_1$ and $V_2$ according to (for details see App.~\ref{App:E symmetry on igSSB})
%
\begin{equation}
    E|_{V_1}=S_{12}\,,\quad E|_{V_2}= S_{21}+U_{2}\,,
\end{equation}
where $S_{12}$ maps $V_1$ to $V_2$ and acts on operators as
\begin{equation}
    S_{12}:\left(\sigma_{j+\half}^z\,, \sigma_{j-\half}^x\sigma_{j+\half}^x\right)\longmapsto \left(\wt{\sigma}^z_{j}\wt{\sigma}^z_{j+1}\,, \wt{\sigma}^x_{j}\right)\,,
\end{equation}
which is precisely the familiar Kramers-Wannier duality map. $S_{21}$ implements the inverse map sending a state $|\vec{p}\,,\vec{q}\rangle \in V_2$ to  $|\vec{0}\,,\vec{q}\rangle \in V_1$ while $U_2$ is a $\Z_2$ symmetry operation that acts within $V_2$ as $p_j\to -p_j \ \text{mod 3}$, which may be understood as the symmetry dual (under gauging) to $U$. These satisfy the following operator relations
\begin{equation}
    S_{21}S_{12}=1+U\,, \qquad S_{12}S_{21}=1+U_{2}\,. 
\end{equation}
Enforced by the $E$ symmetry action, the Hamiltonians \eqref{eq:V1_Ham} and \eqref{eq:V2_Ham} are precisely related by a Kramers-Wannier duality or equivalently a $\Z_2$ gauging.
The transition at $\lambda=1/2$ is in the Ising$\oplus$Ising universality class. Note that the degeneracy between the two gapless Ising states can only be explained by breaking of the non-invertible symmetry $E$, and hence is beyond the standard Landau paradigm.

$O_q$ becomes the spin operator in the first copy of Ising (i.e., in $V_1$), while it vanishes in $V_2$. Meanwhile, $O_p$ becomes the identity in $V_1$ and $\exp\{2\pi i \wt{\sigma}^z/3\}$ in $V_2$.
$\lambda<1/2$ is the $\Rep(S_3)/\Z_2$ SSB phase with three ground states on which $U$ acts trivially, while $\lambda>1/2$ is the $\Rep(S_3)$ SSB phase with three ground states, two of which are the $U$ breaking ferromagnetic states in $V_1$ and the third is the $U$ invariant ground state in $V_2$.

\smallskip\ni\textbf{Symmetry Protected Criticality.} The $\text{Ising}\oplus\text{Ising}$ transition described above lies in a gapless phase exhibiting symmetry protected criticality \cite{scaffidi2017gapless, Verresen:2019igf, Thorngren:2020wet,  Wen:2022tkg, Li:2023knf, Li:2022jbf, Bhardwaj:2024qrf}. Any $\Rep(S_3)$ symmetric deformation of a gapless system lying in this phase can only trigger renormalization group flows that lead to infrared phases of the form $\text{T}\oplus\text{T}/\Z_2$, where T is a $\Z_2$ symmetric theory and $\text{T}/\Z_2$ is the theory obtained after gauging this $\Z_2$ symmetry. The only possible gapped deformations are obtained by choosing T to be a $\Z_2$ symmetric phase. Choosing T to be paramangetic or ferrmagnetic ($\Z_2$ SSB) phase leads respectively to $\Rep(S_3)/\Z_2$ and $\Rep(S_3)$ SSB phases. Consequently, either the system remains gapless with two gapless states, or it becomes gapped with three gapped states. Thus, it is not possible to break criticality without inducing additional order, making this into an intrinsically gapless SSB (igSSB) phase introduced in \cite{Bhardwaj:2024qrf} \footnote{These are analogues of intrinsically gapless SPT (igSPT) phases where there is a single gapless state and it is not possible to gap it without inducing order, i.e. every gapped deformation has multiple degenerate gapped ground states.}.

\smallskip\ni\textbf{Including Impurities.} So far, we have focused on clean systems with translationally invariant Hamiltonians. In a system with a global symmetry, it is natural to also consider the symmetry twisted sectors. The twisted sector Hilbert space for group like, i.e.\ invertible, symmetry operators is isomorphic to the untwisted sector Hilbert space; however, some Hamiltonian terms located at the symmetry twist are modified. E.g. inserting a $U$-twist  at site $j_0$ on the lattice corresponds to the modification
\begin{equation}
    \sigma^x_{j_0-\half}\sigma^x_{j_0+\half} \longrightarrow -    \sigma^x_{j_0-\half}\sigma^x_{j_0+\half}\,, 
\end{equation}
in the Hamiltonian. Note that this only alters the operators $X_{j_0}^{(2I+1)}$ in \eqref{eq:Ham_3site}. In contrast, the twisted sector Hilbert space corresponding to a non-invertible symmetry operation is not isomorphic to the untwisted sector Hilbert space. Inserting an $E$ symmetry twist at site $j_0$ corresponds to modifying the Hilbert space by inserting an impurity with an associated two-dimensional state space at $j_0$. We denote the Pauli operators acting on the impurity Hilbert space as $\sigma^{\mu}_{j_0}$. Then the corresponding $E$-twisted Hamiltonian is given by making the following modifications in \eqref{eq:Ham_3site} 
\begin{equation}
    \hspace{-7pt}Z_{j_0}\left(Z_{j_0+1}\right)^\alpha \to  Z_{j_0}\omega^{\sigma^{z}_{j_0}}\left(Z_{j_0+1}\right)^\alpha\,,
    \ 
    \Gamma_{j_0} \to    \Gamma_{j_0}\sigma^{x}_{j_0}\,,
\end{equation}
where $\alpha=\pm 1$. These symmetry defects satisfy the $\Rep(S_3)$  multiplication rules in eq.~\eqref{fusion}. The Hamiltonian with two $U$ symmetry twists, at the sites $j_0$ and $j_0+n$ is unitarily mapped to the untwisted Hamiltonian by conjugating with $\prod_{\ell =0}^{n-1}\sigma^{z}_{j_0+\half+\ell}$. Similarly,the Hamiltonian with a $U$ twist at $j_0$ and an $E$ twist at $j_0+n$ can be mapped to a Hamiltonian with a single $E$ symmetry twist at $j_0+n$ by conjugating with the unitary $\prod_{\ell =0}^{n-1}\sigma^{z}_{j_0+\half+\ell}\sigma_{j_0+n}$. Finally, let us describe the multiplication rules of $E$ symmetry twists. Consider the case where both symmetry twists/impurities are located at the same site $j_0$\footnote{The more general case requires defining an isomorphism between two Hilbert spaces related by translating the impurity by a single lattice spacing \cite{Bhardwaj:2024kvy}.}. Denote the Pauli operators on the two impurity Hilbert spaces as $\sigma^{\mu}_{j_0}$ and $\wt{\sigma}^{\mu}_{j_0}$ respectively. Then the corresponding $E\otimes E$-twisted Hamiltonian is given by making the following modifications in \eqref{eq:Ham_3site} 
\begin{equation}
    Z_{j_0}\left(Z_{j_0+1}\right)^\alpha \to  Z_{j_0}\omega^{\sigma^{z}_{j_0}+\wt{\sigma}^{z}_{j_0}}\left(Z_{j_0+1}\right)^\alpha\,, \ \, 
    \Gamma_{j_0} \to    \Gamma_{j_0}\sigma^{x}_{j_0}\wt{\sigma}^{x}_{j_0}\,.
\end{equation}  
The four dimensional impurity state space decomposes as a sum of symmetry twisted sectors as
\begin{equation}
    (\bC^4)_{E\otimes E}=     (\bC^2)_E\oplus (\bC)_1\oplus (\bC)_{U}\,,
\end{equation}
where $ (\bC^2)_E= \{ |0\rangle|0\rangle, |1\rangle|1\rangle \}_\C$ while $(\bC)_1$ and $(\bC)_U$ are spanned by $|0\rangle|1\rangle + |1\rangle|0\rangle$ and $|0\rangle|1\rangle - |1\rangle|0\rangle$ respectively.

\smallskip
\ni{\bf Outlook.}
In a companion paper \cite{Bhardwaj:2024kvy}, we discuss more systematically (1+1)d lattice models \cite{Feiguin:2006ydp, Aasen:2016dop, Aasen:2020jwb, Lootens:2021tet, Lootens:2022avn, Inamura:2021szw} with fusion category symmetries, using the anyon chain and constructing gapped and gapless phases. This showcases an example of more general constructions of lattice models with non-standard symmetries, which host novel phases and phase-transitions. 
This can be extended to higher-dimensions using the lattice models in \cite{Delcamp:2023kew, Inamura:2023qzl}, opening up new avenues to study phases in $d>2$, with potential applications in the study of quantum magnets, which have rich phase diagrams realizing beyond-Landau spin liquid phases.






\newpage
\onecolumngrid

\bigskip

\noindent
\textbf{Acknowledgements.}
We thank \"Omer  Aksoy, Arkya Chatterjee, Luisa Eck, Paul Fendley, Sanjay Moudgalya, Klaus M{\o}lmer and Xiao-Gang Wen for discussions.  We thank \"Omer  Aksoy, Arkya Chatterjee, and Xiao-Gang Wen for coordinating submission of their related work \cite{Chatterjee:2024ych} with ours. 
LB thanks Neils Bohr International Academy for hospitality, where a part of this work was completed. 
LB is funded as a Royal Society University Research Fellow through grant
URF{\textbackslash}R1\textbackslash231467. The work of SSN is supported by the UKRI Frontier Research Grant, underwriting the ERC Advanced Grant "Generalized Symmetries in Quantum Field Theory and Quantum Gravity” and the Simons Foundation Collaboration on ``Special Holonomy in Geometry, Analysis, and Physics", Award ID: 724073, Schafer-Nameki. The work of AT is funded by Villum Fonden Grant no. VIL60714.


\appendix

\section{Ground States of the Hamiltonian $H_4$}\label{H4_ground_states}
In this section we provide more details for the derivation of the ground states of the Hamiltonian
\begin{equation}\label{app:H4}
    H_4 =-\half\sum_j \left(1+\sigma^{x}_{j-\half}\Gamma_j\sigma^x_{j+\half}\right) -\frac{1}{6}\sum_{j}\sum_{\alpha=\pm 1}\left[ 1+\alpha\sigma^{z}_{j+\half}\right]\left[\sum_{n=0}^{2} Z^n_{j}Z_{j+1}^{-\alpha n}\right]    \,.
\end{equation}
The first and second terms commute, so they can be diagonalized simultaneously. We first consider the $+1$ eigenspace of the second term in \eqref{app:H4}. Notice that this enforces a relation between the variables $p_{j-1}$, $q_{j+\frac{1}{2}}$ and $p_{j+1}$, namely 
\begin{equation}
\begin{cases}
    p_{j+1} - p_j = 0 \;  \quad \text{if} \quad q_{j+\frac{1}{2}} = 0 \\
    p_{j+1} + p_j = 0 \;(\text{mod } 3) \quad \text{if} \quad q_{j+\frac{1}{2}} = 1 \,.
\end{cases}   
\end{equation}
We first consider states $| \vec p,\vec q \rangle$ such that $p_j \neq 0$ and $p_j + p_{j-1} = q_{j-\frac{1}{2}}$ (mod 2). It is easy to check that all of these states form a single orbit under the action of the first term in \eqref{app:H4}. Therefore we can identify a first ground state of the Hamiltonian as
\begin{equation}
{\sum_{\vec{p}\,,\vec{q}}}' |\vec{p}\,,\vec{q}\rangle \quad,\quad p_j \neq 0 \;, p_j + p_{j-1} = q_{j-\frac{1}{2}} \text{ mod } 2
\end{equation} 
with the appropriate normalization, where $\sum_{\vec{p}\,,\vec{q}}'$ denotes a sum restricted precisely to states satisfying the above condition. 
We can then consider states that have $p_j=0$ for every site $j$. Among these, all the $|\vec 0\,,\vec{q}\rangle$ states such that $\sum_j q_{j+\frac{1}{2}} = 0$ (mod 2) are permuted among each other by the first term in the Hamiltonian \eqref{app:H4}. Similarly, the states such that $\sum_j q_{j+\frac{1}{2}} = 1$ (mod 2) form an orbit under the action of the first term. We can then define two eigenstates of \eqref{app:H4} as 
\begin{equation}
\begin{aligned}
    &|+\rangle \sim  \sum_{\vec q} |\vec 0, \vec q \rangle \quad,\quad \sum_j q_{j+\frac{1}{2}} = 0 \text{ mod }2 \\
    &|-\rangle \sim  \sum_{\vec q} |\vec 0, \vec q \rangle \quad,\quad \sum_j q_{j+\frac{1}{2}} = 1 \text{ mod }2 \,.
\end{aligned}    
\end{equation}
with the appropriate normalization. Notice that $|+ \rangle$ and $|- \rangle$ are respectively even and odd under the $\Z_2$ symmetry generated by 
\begin{equation}
    U=\prod_{j}\sigma^z_{j+\half}\,.
\end{equation}
We therefore observe that the $\Z_2$ symmetry must be spontaneously broken in the gapped phase realized by the ground states of this Hamiltonian. In this instance it is well-known (see for example \cite{Bhardwaj:2023idu}) that the ground states are given by the two combinations 
\begin{equation}
    \frac{|+\rangle \pm |-\rangle}{2} \,.
\end{equation}
Therefore we find the other two ground states of $H_4$ as 
\begin{equation}
\begin{split}
&\sum_{\vec{q}}|\vec{0}\,,\vec{q}\rangle\,,  \\
&\sum_{\vec{q}}(-1)^{\sum_jq_{j+\half}}|\vec{0}\,,\vec{q}\rangle\,,
\end{split}
\end{equation}
with the appropriate normalization. 
This reproduces all the ground states listed in \eqref{eq:GS_4}.

\section{Proof that $\Rep(S_3)$ Symmetries Commute with the Hamiltonian}\label{commute}
The operator $U$, defined in \eqref{U}, which measures the total spin parity of the qubits, clearly commutes with the Hamiltonian \eqref{eq:Ham_3site}.
Let us then consider the symmetry operator $E \in \Rep(S_3)$, which we defined in \eqref{RepS3_operators}, and show that it also commutes with $H$. For convenience, let us define the operators 
\begin{equation}
	Q_j^\pm = \frac{1}{2} \left( 1 \pm \prod_{i=0}^{j-1} \sigma_{i+\frac{1}{2}}^z\right) \,,
\end{equation}
so that we can rewrite $T_s$ in \eqref{RepS3_operators} as
\begin{equation}
	T_s = \prod_{j=1}^L \left[ Q_j^+ X_j^s + Q_j^- X_j^{2s} \right] \,.
\end{equation}
Let us start by considering the term 
\begin{equation}
	X_j^{(2I+s)} = (X_j)^I (\sigma^x_{j-\frac{1}{2}} \Gamma_j \sigma^x_{j+\frac{1}{2}} )^s
\end{equation}
in the Hamiltonian. 
The $X_j$ operators clearly commute with each other, so we only need to care about the possible non-commutativity of $\sigma^z$  with  $\sigma^x$, as well as $\Gamma$ and $X^s$, when the two operators are at the same lattice site.  In particular, the terms in $T_s$ that overlap on the lattice with 
\begin{equation}\label{app:sigma_x}
	\sigma^x_{j-\frac{1}{2}} \Gamma_j \sigma^x_{j+\frac{1}{2}}
\end{equation}
are
\begin{equation}\label{app:T_comm}
	\left( Q_j^+ X_j^s + Q_j^- X_j^{2s} \right)  \left( Q_{j+1}^+ X_{j+1}^s + Q_{j+1}^- X_{j+1}^{2s} \right) \dots \,.
\end{equation}
First of all, it is easy to show that the $\left( Q_{k}^+ X_{k}^s + Q_{k}^- X_{k}^{2s} \right)$ operators commute with \eqref{app:sigma_x} when $k \geq j+1$. This is because in this instance there is no overlap between $\Gamma_j$, which acts as $1 \leftrightarrow 2$ on the $p_j$ variable, and $X_k^{s,2s}$, which act on $p_k$. Moreover, in this case every $Q_{k}^\pm$ contains both $\sigma^z_{j+\frac{1}{2}}$ and $\sigma^z_{j-\frac{1}{2}}$, which implies its commutation with a simultaneous flip of the $q_{j+\frac{1}{2}}$ and  $q_{j-\frac{1}{2}}$ variables due to the action of  \eqref{app:sigma_x}. Therefore, we only need to worry about the first term in \eqref{app:T_comm}, which involves $X_j^{s,2s}$ and $Q_j^\pm$. Notice that since 
\begin{equation}
    Q_j^\pm = 1 \pm \sigma_{\frac{1}{2}}^z \dots \sigma_{j-\frac{1}{2}}^z \,,
\end{equation}
acting first on a state $|\vec p,\vec q \rangle$ with \eqref{app:sigma_x} effectively changes $Q_j^+ \rightarrow Q_j^-$. The fact that the full operator $T_s$ commutes with this term follows then since, as one can check,
\begin{equation}
    \Gamma_j X_j |p_j \rangle = X_j^{2} \Gamma_j |p_j \rangle \quad,\quad \Gamma_j X_j^2 |p_j \rangle = X_j \Gamma_j |p_j \rangle \,.
\end{equation}
Now let us consider the second term in the Hamiltonian \eqref{eq:Ham_3site}, namely 
\begin{equation}
    P_{j+\frac{1}{2}}^{(2I+s)} = \frac{1}{6} \left[ 1+(-1)^s \sigma^z_{j+\frac{1}{2}}\right] \left[ \sum_{n=0}^2 \omega^{-In}Z_j^n Z_{j+1}^{(2s-1)n}\right] \,.
\end{equation}
Let us focus on $P_{j+\frac{1}{2}}^{(0)}$ and $ P_{j+\frac{1}{2}}^{(1)}$, with the other terms being analogous. We first observe that considering the $+1$ eigenspace of these operators enforces a relation between the variables $p_{j-1}$, $q_{j+\frac{1}{2}}$ and $p_{j+1}$, namely 
\begin{equation}
\begin{cases}
    p_{j+1} - p_j = 0 \;  \quad \text{if} \quad q_{j+\frac{1}{2}} = 0 \\
    p_{j+1} + p_j = 0 \;(\text{mod } 3) \quad \text{if} \quad q_{j+\frac{1}{2}} = 1 \,.
\end{cases}   
\end{equation}
To show that $E$ commutes with $P_{j+\frac{1}{2}}$, we essentially need to show that its action respects these relations. Let us consider e.g.\ $T_1$ for concreteness. The relevant terms are
\begin{equation}\label{app:T_comm2}
   (Q_j^+ X_j + Q_j^- X_j^2)(Q_{j+1}^+ X_{j+1} +Q_{j+1}^- X_{j+1}^2) \,.
\end{equation}
We first consider the case $q_{j+\frac{1}{2}}=0$. The first term in \eqref{app:T_comm2} acts on a state $|\vec p,\vec q \rangle$ as $p_{j} \rightarrow p_j+ k$, with $k=1$ if $Q_j^+=+1$ and $k=2$ if $Q_j^- = 1$. Moreover, notice that since $q_{j+1} = 0$, the total parity evaluated by $Q_{j+1}^\pm$ does not change and therefore also the second term acts by exactly the same shift $p_{j+1} \rightarrow p_{j+1} + k$. Therefore the relation $p_{j+1}- p_j = 0$ is preserved. Now let us consider the case $q_{j+\frac{1}{2}} = 1$. Again the term in the first parenthesis in \eqref{app:T_comm2} acts as $p_j \rightarrow p_j + k$ depending if either $Q_j^+$ or $Q_j^-$ is non-zero. However, since $q_{j+\frac{1}{2}} = 1$, now $Q_{j+1}^\pm$ detects precisely the opposite parity. Therefore the second term now acts as $p_{j+1} \rightarrow p_{j+1} + 2k \; (\text{mod }3)$. This means that also in this case the relation $p_{j+1} + p_j = 0 \;(\text{mod } 3)$ is preserved. Therefore $T_1$ commutes with $P_{j+\frac{1}{2}}^{(0)}, \; P_{j+\frac{1}{2}}^{(1)}$. The other cases are completely analogous.


\section{Action of $E$ in the igSSB-Phase}
\label{App:E symmetry on igSSB}
Let us consider how the symmetry $E\in \Rep(S_3)$ acts on the model $H_{3,4}(\lambda)$ describing the transition between the $\Rep(S_3)$ SSB and $\Rep(S_3)/\Z_2$ SSB gapped phases. As described in the main text, the low lying spectrum within this phase contains two dynamically disconnected state spaces $V_1$ and $V_2$. The state space $V_{1}$ spanned by states of the form $|\vec{0},\vec{q}\rangle$ can further be decomposed as 
\begin{equation}
    V_{1}=V_{1}^{+}\oplus V_{1}^{-}\,,
\end{equation}
where $V_1^s$ is the space with a $s$ eigenvalue under $U$. More specifically, the states with $\sum_{j}q_{j+\half}=0\text{ mod 2}$ lie in $V_{1}^{+}$, while the states $\sum_{j}q_{j+\half}=1\text{ mod 2}$ lie in $V_{1}^{-}$. It can be seen from  Eq.~\eqref{RepS3_operators} that $V_{1}^{-}$ is in the kernel of the $E$ symmetry operator, while a state in $V_{1}^{+}$ transforms as
\begin{equation}
    E|\vec{0},\vec{q}\rangle = |\vec{p}_1(\vec{q}),\vec{q}\rangle + 
    |\vec{p}_2(\vec{q}),\vec{q}\rangle  \quad \in  \ V_2\,,
    \label{eq:E_on_V1}
\end{equation}
where $\vec{p}_{1}(\vec{q})$ and $\vec{p}_{2}(\vec{q})$ are two configurations of the qutrit degrees of freedom such that $p_{j}\neq 0 $ and $q_{j+1/2}=p_{j+1}+p_j \ \text{mod 2}$. Since $E$ maps states in $V_1$ to states completely lying within $V_2$, we denote
\begin{equation}
    E\Big|_{V_1}=S_{12}\,,
\end{equation}
The map $S_{12}$ sends a state with $q_{j+\half}= 0$, i.e.\ $\sigma^{z}_{j+\half}= 1$, to the sum of states with $p_{j}=p_{j+1}$, i.e.\ $\wt{\sigma}^z_j\wt{\sigma}^z_{j+1}=1$. Similarly, $S_{12}$ sends a state with $q_{j+\half}= 1$, i.e.\ $\sigma^{z}_{j+\half}= -1$, to the sum of states with $p_{j}=-p_{j+1}$, i.e.\ $\wt{\sigma}^z_j\wt{\sigma}^z_{j+1}=-1$. 
Similarly one can verify that $S_{12}$ maps $\sigma^{x}_{j-\half}\sigma^{x}_{j+\half}$ to $\wt{\sigma}^x_{j}$.
To summarize $S_{12}$ implements the familiar Kramers-Wannier map on operators
\begin{equation}
    S_{12}:\left(\sigma_{j+\half}^z\,, \sigma_{j-\half}^x\sigma_{j+\half}^x\right)\longmapsto \left(\wt{\sigma}^z_{j}\wt{\sigma}^z_{j+1}\,, \wt{\sigma}^x_{j}\right)\,.
\end{equation}
Note that $\vec{p}_{1}(\vec{q})$ and $\vec{p}_{2}(\vec{q})$ in eq.~\eqref{eq:E_on_V1} are related by inverting all $p_{j}$ to $-p_{j}$. We may therefore define a $\Z_2$  operation $U_2=\prod_j\Gamma_j$ acting within $V_2$ that implements
\begin{equation}
U_2 |\vec{p}\,,\vec{q}\rangle=|-\vec{p}\,,\vec{q}\rangle\,.
\end{equation}   
It then follows from Eq.~\eqref{RepS3_operators} that a state in $V_2$ transforms under $E$ as
\begin{equation}
    E|\vec{p}\,,\vec{q}\rangle= |\vec{0}\,,\vec{q}\rangle+U_2|\vec{p}\,,\vec{q}\rangle =: S_{21}|\vec{p}\,,\vec{q}\rangle+ U_{2}|\vec{p}\,,\vec{q}\rangle\,.
\end{equation}
To summarize, the $E$ symmetry is realized as
\begin{equation}
    E\Big|_{V_1\oplus V_2}=S_{12}+S_{21}+U_2\,,
\end{equation}
where $S_{12}$ and $S_{21}$ satisfy the relations 
\begin{equation}
    S_{21}S_{12}=1+U\,, \qquad S_{12}S_{21}=1+U_{2}\,. 
\end{equation}


\twocolumngrid
\bibliography{GenSym}

\begin{thebibliography}{47}%
\makeatletter
\providecommand \@ifxundefined [1]{%
 \@ifx{#1\undefined}
}%
\providecommand \@ifnum [1]{%
 \ifnum #1\expandafter \@firstoftwo
 \else \expandafter \@secondoftwo
 \fi
}%
\providecommand \@ifx [1]{%
 \ifx #1\expandafter \@firstoftwo
 \else \expandafter \@secondoftwo
 \fi
}%
\providecommand \natexlab [1]{#1}%
\providecommand \enquote  [1]{``#1''}%
\providecommand \bibnamefont  [1]{#1}%
\providecommand \bibfnamefont [1]{#1}%
\providecommand \citenamefont [1]{#1}%
\providecommand \href@noop [0]{\@secondoftwo}%
\providecommand \href [0]{\begingroup \@sanitize@url \@href}%
\providecommand \@href[1]{\@@startlink{#1}\@@href}%
\providecommand \@@href[1]{\endgroup#1\@@endlink}%
\providecommand \@sanitize@url [0]{\catcode `\\12\catcode `\$12\catcode
  `\&12\catcode `\#12\catcode `\^12\catcode `\_12\catcode `\%12\relax}%
\providecommand \@@startlink[1]{}%
\providecommand \@@endlink[0]{}%
\providecommand \url  [0]{\begingroup\@sanitize@url \@url }%
\providecommand \@url [1]{\endgroup\@href {#1}{\urlprefix }}%
\providecommand \urlprefix  [0]{URL }%
\providecommand \Eprint [0]{\href }%
\providecommand \doibase [0]{http://dx.doi.org/}%
\providecommand \selectlanguage [0]{\@gobble}%
\providecommand \bibinfo  [0]{\@secondoftwo}%
\providecommand \bibfield  [0]{\@secondoftwo}%
\providecommand \translation [1]{[#1]}%
\providecommand \BibitemOpen [0]{}%
\providecommand \bibitemStop [0]{}%
\providecommand \bibitemNoStop [0]{.\EOS\space}%
\providecommand \EOS [0]{\spacefactor3000\relax}%
\providecommand \BibitemShut  [1]{\csname bibitem#1\endcsname}%
\let\auto@bib@innerbib\@empty
\bibitem [{\citenamefont {Gaiotto}\ \emph {et~al.}(2015)\citenamefont
  {Gaiotto}, \citenamefont {Kapustin}, \citenamefont {Seiberg},\ and\
  \citenamefont {Willett}}]{Gaiotto:2014kfa}%
  \BibitemOpen
  \bibfield  {author} {\bibinfo {author} {\bibfnamefont {D.}~\bibnamefont
  {Gaiotto}}, \bibinfo {author} {\bibfnamefont {A.}~\bibnamefont {Kapustin}},
  \bibinfo {author} {\bibfnamefont {N.}~\bibnamefont {Seiberg}}, \ and\
  \bibinfo {author} {\bibfnamefont {B.}~\bibnamefont {Willett}},\ }\href
  {\doibase 10.1007/JHEP02(2015)172} {\bibfield  {journal} {\bibinfo  {journal}
  {JHEP}\ }\textbf {\bibinfo {volume} {02}},\ \bibinfo {pages} {172} (\bibinfo
  {year} {2015})},\ \Eprint {http://arxiv.org/abs/1412.5148} {arXiv:1412.5148
  [hep-th]} \BibitemShut {NoStop}%
\bibitem [{Note1()}]{Note1}%
  \BibitemOpen
  \bibinfo {note} {See \cite {Schafer-Nameki:2023jdn, Shao:2023gho} for recent
  reviews on the topic.}\BibitemShut {Stop}%
\bibitem [{\citenamefont {Bhardwaj}\ \emph
  {et~al.}(2023{\natexlab{a}})\citenamefont {Bhardwaj}, \citenamefont
  {Bottini}, \citenamefont {Pajer},\ and\ \citenamefont
  {Schafer-Nameki}}]{Bhardwaj:2023fca}%
  \BibitemOpen
  \bibfield  {author} {\bibinfo {author} {\bibfnamefont {L.}~\bibnamefont
  {Bhardwaj}}, \bibinfo {author} {\bibfnamefont {L.~E.}\ \bibnamefont
  {Bottini}}, \bibinfo {author} {\bibfnamefont {D.}~\bibnamefont {Pajer}}, \
  and\ \bibinfo {author} {\bibfnamefont {S.}~\bibnamefont {Schafer-Nameki}},\
  }\href@noop {} {\  (\bibinfo {year} {2023}{\natexlab{a}})},\ \Eprint
  {http://arxiv.org/abs/2310.03786} {arXiv:2310.03786 [cond-mat.str-el]}
  \BibitemShut {NoStop}%
\bibitem [{\citenamefont {Bhardwaj}\ \emph
  {et~al.}(2023{\natexlab{b}})\citenamefont {Bhardwaj}, \citenamefont
  {Bottini}, \citenamefont {Pajer},\ and\ \citenamefont
  {Schafer-Nameki}}]{Bhardwaj:2023idu}%
  \BibitemOpen
  \bibfield  {author} {\bibinfo {author} {\bibfnamefont {L.}~\bibnamefont
  {Bhardwaj}}, \bibinfo {author} {\bibfnamefont {L.~E.}\ \bibnamefont
  {Bottini}}, \bibinfo {author} {\bibfnamefont {D.}~\bibnamefont {Pajer}}, \
  and\ \bibinfo {author} {\bibfnamefont {S.}~\bibnamefont {Schafer-Nameki}},\
  }\href@noop {} {\  (\bibinfo {year} {2023}{\natexlab{b}})},\ \Eprint
  {http://arxiv.org/abs/2310.03784} {arXiv:2310.03784 [hep-th]} \BibitemShut
  {NoStop}%
\bibitem [{\citenamefont {Bhardwaj}\ \emph
  {et~al.}(2023{\natexlab{c}})\citenamefont {Bhardwaj}, \citenamefont
  {Bottini}, \citenamefont {Pajer},\ and\ \citenamefont
  {Schafer-Nameki}}]{Bhardwaj:2023bbf}%
  \BibitemOpen
  \bibfield  {author} {\bibinfo {author} {\bibfnamefont {L.}~\bibnamefont
  {Bhardwaj}}, \bibinfo {author} {\bibfnamefont {L.~E.}\ \bibnamefont
  {Bottini}}, \bibinfo {author} {\bibfnamefont {D.}~\bibnamefont {Pajer}}, \
  and\ \bibinfo {author} {\bibfnamefont {S.}~\bibnamefont {Schafer-Nameki}},\
  }\href@noop {} {\  (\bibinfo {year} {2023}{\natexlab{c}})},\ \Eprint
  {http://arxiv.org/abs/2312.17322} {arXiv:2312.17322 [hep-th]} \BibitemShut
  {NoStop}%
\bibitem [{\citenamefont {Bhardwaj}\ \emph
  {et~al.}(2024{\natexlab{a}})\citenamefont {Bhardwaj}, \citenamefont {Pajer},
  \citenamefont {Schafer-Nameki},\ and\ \citenamefont
  {Warman}}]{Bhardwaj:2024qrf}%
  \BibitemOpen
  \bibfield  {author} {\bibinfo {author} {\bibfnamefont {L.}~\bibnamefont
  {Bhardwaj}}, \bibinfo {author} {\bibfnamefont {D.}~\bibnamefont {Pajer}},
  \bibinfo {author} {\bibfnamefont {S.}~\bibnamefont {Schafer-Nameki}}, \ and\
  \bibinfo {author} {\bibfnamefont {A.}~\bibnamefont {Warman}},\ }\href@noop {}
  {\  (\bibinfo {year} {2024}{\natexlab{a}})},\ \Eprint
  {http://arxiv.org/abs/2403.00905} {arXiv:2403.00905 [cond-mat.str-el]}
  \BibitemShut {NoStop}%
\bibitem [{Note2()}]{Note2}%
  \BibitemOpen
  \bibinfo {note} {For related works using similar SymTFT approaches for
  invertible symmetries to determine gapped phases, see \cite {Vanhove:2018wlb,
  Chatterjee:2022jll, Chatterjee:2022tyg, Wen:2023otf, Huang:2023pyk,
  Moradi:2022lqp, Chen:2022wvy}. For gapped phases with non-invertible
  symmetries in lattice models see \cite
  {Fechisin:2023dkj,Seifnashri:2024dsd}.}\BibitemShut {Stop}%
\bibitem [{\citenamefont {Ji}\ and\ \citenamefont {Wen}(2020)}]{Ji:2019jhk}%
  \BibitemOpen
  \bibfield  {author} {\bibinfo {author} {\bibfnamefont {W.}~\bibnamefont
  {Ji}}\ and\ \bibinfo {author} {\bibfnamefont {X.-G.}\ \bibnamefont {Wen}},\
  }\href {\doibase 10.1103/PhysRevResearch.2.033417} {\bibfield  {journal}
  {\bibinfo  {journal} {Phys. Rev. Res.}\ }\textbf {\bibinfo {volume} {2}},\
  \bibinfo {pages} {033417} (\bibinfo {year} {2020})},\ \Eprint
  {http://arxiv.org/abs/1912.13492} {arXiv:1912.13492 [cond-mat.str-el]}
  \BibitemShut {NoStop}%
\bibitem [{\citenamefont {Gaiotto}\ and\ \citenamefont
  {Kulp}(2021)}]{Gaiotto:2020iye}%
  \BibitemOpen
  \bibfield  {author} {\bibinfo {author} {\bibfnamefont {D.}~\bibnamefont
  {Gaiotto}}\ and\ \bibinfo {author} {\bibfnamefont {J.}~\bibnamefont {Kulp}},\
  }\href {\doibase 10.1007/JHEP02(2021)132} {\bibfield  {journal} {\bibinfo
  {journal} {JHEP}\ }\textbf {\bibinfo {volume} {02}},\ \bibinfo {pages} {132}
  (\bibinfo {year} {2021})},\ \Eprint {http://arxiv.org/abs/2008.05960}
  {arXiv:2008.05960 [hep-th]} \BibitemShut {NoStop}%
\bibitem [{\citenamefont {Apruzzi}\ \emph {et~al.}(2021)\citenamefont
  {Apruzzi}, \citenamefont {Bonetti}, \citenamefont {Etxebarria}, \citenamefont
  {Hosseini},\ and\ \citenamefont {Schafer-Nameki}}]{Apruzzi:2021nmk}%
  \BibitemOpen
  \bibfield  {author} {\bibinfo {author} {\bibfnamefont {F.}~\bibnamefont
  {Apruzzi}}, \bibinfo {author} {\bibfnamefont {F.}~\bibnamefont {Bonetti}},
  \bibinfo {author} {\bibfnamefont {I.~n.~G.}\ \bibnamefont {Etxebarria}},
  \bibinfo {author} {\bibfnamefont {S.~S.}\ \bibnamefont {Hosseini}}, \ and\
  \bibinfo {author} {\bibfnamefont {S.}~\bibnamefont {Schafer-Nameki}},\
  }\href@noop {} {\  (\bibinfo {year} {2021})},\ \Eprint
  {http://arxiv.org/abs/2112.02092} {arXiv:2112.02092 [hep-th]} \BibitemShut
  {NoStop}%
\bibitem [{\citenamefont {Freed}\ \emph {et~al.}(2022)\citenamefont {Freed},
  \citenamefont {Moore},\ and\ \citenamefont {Teleman}}]{Freed:2022qnc}%
  \BibitemOpen
  \bibfield  {author} {\bibinfo {author} {\bibfnamefont {D.~S.}\ \bibnamefont
  {Freed}}, \bibinfo {author} {\bibfnamefont {G.~W.}\ \bibnamefont {Moore}}, \
  and\ \bibinfo {author} {\bibfnamefont {C.}~\bibnamefont {Teleman}},\
  }\href@noop {} {\  (\bibinfo {year} {2022})},\ \Eprint
  {http://arxiv.org/abs/2209.07471} {arXiv:2209.07471 [hep-th]} \BibitemShut
  {NoStop}%
\bibitem [{\citenamefont {Bhardwaj}\ and\ \citenamefont
  {Schafer-Nameki}(2023{\natexlab{a}})}]{Bhardwaj:2023ayw}%
  \BibitemOpen
  \bibfield  {author} {\bibinfo {author} {\bibfnamefont {L.}~\bibnamefont
  {Bhardwaj}}\ and\ \bibinfo {author} {\bibfnamefont {S.}~\bibnamefont
  {Schafer-Nameki}},\ }\href@noop {} {\  (\bibinfo {year}
  {2023}{\natexlab{a}})},\ \Eprint {http://arxiv.org/abs/2305.17159}
  {arXiv:2305.17159 [hep-th]} \BibitemShut {NoStop}%
\bibitem [{\citenamefont {Eck}\ and\ \citenamefont
  {Fendley}(2023{\natexlab{a}})}]{Eck:2023gic}%
  \BibitemOpen
  \bibfield  {author} {\bibinfo {author} {\bibfnamefont {L.}~\bibnamefont
  {Eck}}\ and\ \bibinfo {author} {\bibfnamefont {P.}~\bibnamefont {Fendley}},\
  }\href@noop {} {\  (\bibinfo {year} {2023}{\natexlab{a}})},\ \Eprint
  {http://arxiv.org/abs/2302.14081} {arXiv:2302.14081 [cond-mat.stat-mech]}
  \BibitemShut {NoStop}%
\bibitem [{\citenamefont {Eck}\ and\ \citenamefont
  {Fendley}(2023{\natexlab{b}})}]{eck2023critical}%
  \BibitemOpen
  \bibfield  {author} {\bibinfo {author} {\bibfnamefont {L.}~\bibnamefont
  {Eck}}\ and\ \bibinfo {author} {\bibfnamefont {P.}~\bibnamefont {Fendley}},\
  }\href@noop {} {\bibfield  {journal} {\bibinfo  {journal} {Physical Review
  B}\ }\textbf {\bibinfo {volume} {108}},\ \bibinfo {pages} {125135} (\bibinfo
  {year} {2023}{\natexlab{b}})}\BibitemShut {NoStop}%
\bibitem [{Note3()}]{Note3}%
  \BibitemOpen
  \bibinfo {note} {We have chosen to normalize the terms such that each term is
  separately a projection operator, i.e., it has eigenvalues 0 and
  1.}\BibitemShut {Stop}%
\bibitem [{Note4()}]{Note4}%
  \BibitemOpen
  \bibinfo {note} {Strictly speaking, these two ground states are degenerate
  only in the infinite size/thermodynamic limit as at finite volume, one may
  add $\protect \mathbb {Z}_2$-symmetric terms to the Hamiltonian to create an
  energy gap between symmetric and anti-symmetric combinations of the ground
  states.}\BibitemShut {Stop}%
\bibitem [{Note5()}]{Note5}%
  \BibitemOpen
  \bibinfo {note} {More precisely, the action of $\protect \mathsf {Rep}(S_3)$
  is irreducible on the ground states exhibiting cluster decomposition in the
  infinite volume limit (such ground states are also known as vacua). That is,
  one can generate all vacua starting from any one vacuum and acting on it by
  $\protect \mathsf {Rep}(S_3)$ generators. All the ground states of gapped
  phases displayed in this paper are actually vacuum states.}\BibitemShut
  {Stop}%
\bibitem [{\citenamefont {Bhardwaj}\ and\ \citenamefont
  {Schafer-Nameki}(2023{\natexlab{b}})}]{Bhardwaj:2023wzd}%
  \BibitemOpen
  \bibfield  {author} {\bibinfo {author} {\bibfnamefont {L.}~\bibnamefont
  {Bhardwaj}}\ and\ \bibinfo {author} {\bibfnamefont {S.}~\bibnamefont
  {Schafer-Nameki}},\ }\href@noop {} {\  (\bibinfo {year}
  {2023}{\natexlab{b}})},\ \Eprint {http://arxiv.org/abs/2304.02660}
  {arXiv:2304.02660 [hep-th]} \BibitemShut {NoStop}%
\bibitem [{\citenamefont {Bhardwaj}\ \emph
  {et~al.}(2024{\natexlab{b}})\citenamefont {Bhardwaj}, \citenamefont
  {Bottini}, \citenamefont {Schafer-Nameki},\ and\ \citenamefont
  {Tiwari}}]{Bhardwaj:2024kvy}%
  \BibitemOpen
  \bibfield  {author} {\bibinfo {author} {\bibfnamefont {L.}~\bibnamefont
  {Bhardwaj}}, \bibinfo {author} {\bibfnamefont {L.~E.}\ \bibnamefont
  {Bottini}}, \bibinfo {author} {\bibfnamefont {S.}~\bibnamefont
  {Schafer-Nameki}}, \ and\ \bibinfo {author} {\bibfnamefont {A.}~\bibnamefont
  {Tiwari}},\ }\href@noop {} {\  (\bibinfo {year} {2024}{\natexlab{b}})},\
  \Eprint {http://arxiv.org/abs/2405.05964} {arXiv:2405.05964
  [cond-mat.str-el]} \BibitemShut {NoStop}%
\bibitem [{\citenamefont {Scaffidi}\ \emph {et~al.}(2017)\citenamefont
  {Scaffidi}, \citenamefont {Parker},\ and\ \citenamefont
  {Vasseur}}]{scaffidi2017gapless}%
  \BibitemOpen
  \bibfield  {author} {\bibinfo {author} {\bibfnamefont {T.}~\bibnamefont
  {Scaffidi}}, \bibinfo {author} {\bibfnamefont {D.~E.}\ \bibnamefont
  {Parker}}, \ and\ \bibinfo {author} {\bibfnamefont {R.}~\bibnamefont
  {Vasseur}},\ }\href@noop {} {\bibfield  {journal} {\bibinfo  {journal}
  {Physical Review X}\ }\textbf {\bibinfo {volume} {7}},\ \bibinfo {pages}
  {041048} (\bibinfo {year} {2017})}\BibitemShut {NoStop}%
\bibitem [{\citenamefont {Verresen}\ \emph {et~al.}(2021)\citenamefont
  {Verresen}, \citenamefont {Thorngren}, \citenamefont {Jones},\ and\
  \citenamefont {Pollmann}}]{Verresen:2019igf}%
  \BibitemOpen
  \bibfield  {author} {\bibinfo {author} {\bibfnamefont {R.}~\bibnamefont
  {Verresen}}, \bibinfo {author} {\bibfnamefont {R.}~\bibnamefont {Thorngren}},
  \bibinfo {author} {\bibfnamefont {N.~G.}\ \bibnamefont {Jones}}, \ and\
  \bibinfo {author} {\bibfnamefont {F.}~\bibnamefont {Pollmann}},\ }\href
  {\doibase 10.1103/PhysRevX.11.041059} {\bibfield  {journal} {\bibinfo
  {journal} {Phys. Rev. X}\ }\textbf {\bibinfo {volume} {11}},\ \bibinfo
  {pages} {041059} (\bibinfo {year} {2021})},\ \Eprint
  {http://arxiv.org/abs/1905.06969} {arXiv:1905.06969 [cond-mat.str-el]}
  \BibitemShut {NoStop}%
\bibitem [{\citenamefont {Thorngren}\ \emph {et~al.}(2021)\citenamefont
  {Thorngren}, \citenamefont {Vishwanath},\ and\ \citenamefont
  {Verresen}}]{Thorngren:2020wet}%
  \BibitemOpen
  \bibfield  {author} {\bibinfo {author} {\bibfnamefont {R.}~\bibnamefont
  {Thorngren}}, \bibinfo {author} {\bibfnamefont {A.}~\bibnamefont
  {Vishwanath}}, \ and\ \bibinfo {author} {\bibfnamefont {R.}~\bibnamefont
  {Verresen}},\ }\href {\doibase 10.1103/PhysRevB.104.075132} {\bibfield
  {journal} {\bibinfo  {journal} {Phys. Rev. B}\ }\textbf {\bibinfo {volume}
  {104}},\ \bibinfo {pages} {075132} (\bibinfo {year} {2021})},\ \Eprint
  {http://arxiv.org/abs/2008.06638} {arXiv:2008.06638 [cond-mat.str-el]}
  \BibitemShut {NoStop}%
\bibitem [{\citenamefont {Wen}\ and\ \citenamefont
  {Potter}(2023{\natexlab{a}})}]{Wen:2022tkg}%
  \BibitemOpen
  \bibfield  {author} {\bibinfo {author} {\bibfnamefont {R.}~\bibnamefont
  {Wen}}\ and\ \bibinfo {author} {\bibfnamefont {A.~C.}\ \bibnamefont
  {Potter}},\ }\href {\doibase 10.1103/PhysRevB.107.245127} {\bibfield
  {journal} {\bibinfo  {journal} {Phys. Rev. B}\ }\textbf {\bibinfo {volume}
  {107}},\ \bibinfo {pages} {245127} (\bibinfo {year} {2023}{\natexlab{a}})},\
  \Eprint {http://arxiv.org/abs/2208.09001} {arXiv:2208.09001
  [cond-mat.str-el]} \BibitemShut {NoStop}%
\bibitem [{\citenamefont {Li}\ \emph {et~al.}(2023)\citenamefont {Li},
  \citenamefont {Oshikawa},\ and\ \citenamefont {Zheng}}]{Li:2023knf}%
  \BibitemOpen
  \bibfield  {author} {\bibinfo {author} {\bibfnamefont {L.}~\bibnamefont
  {Li}}, \bibinfo {author} {\bibfnamefont {M.}~\bibnamefont {Oshikawa}}, \ and\
  \bibinfo {author} {\bibfnamefont {Y.}~\bibnamefont {Zheng}},\ }\href@noop {}
  {\  (\bibinfo {year} {2023})},\ \Eprint {http://arxiv.org/abs/2307.04788}
  {arXiv:2307.04788 [cond-mat.str-el]} \BibitemShut {NoStop}%
\bibitem [{\citenamefont {Li}\ \emph {et~al.}(2022)\citenamefont {Li},
  \citenamefont {Oshikawa},\ and\ \citenamefont {Zheng}}]{Li:2022jbf}%
  \BibitemOpen
  \bibfield  {author} {\bibinfo {author} {\bibfnamefont {L.}~\bibnamefont
  {Li}}, \bibinfo {author} {\bibfnamefont {M.}~\bibnamefont {Oshikawa}}, \ and\
  \bibinfo {author} {\bibfnamefont {Y.}~\bibnamefont {Zheng}},\ }\href@noop {}
  {\  (\bibinfo {year} {2022})},\ \Eprint {http://arxiv.org/abs/2204.03131}
  {arXiv:2204.03131 [cond-mat.str-el]} \BibitemShut {NoStop}%
\bibitem [{Note6()}]{Note6}%
  \BibitemOpen
  \bibinfo {note} {These are analogues of intrinsically gapless SPT (igSPT)
  phases where there is a single gapless state and it is not possible to gap it
  without inducing order, i.e. every gapped deformation has multiple degenerate
  gapped ground states.}\BibitemShut {Stop}%
\bibitem [{Note7()}]{Note7}%
  \BibitemOpen
  \bibinfo {note} {The more general case requires defining an isomorphism
  between two Hilbert spaces related by translating the impurity by a single
  lattice spacing \cite {Bhardwaj:2024kvy}.}\BibitemShut {Stop}%
\bibitem [{\citenamefont {Feiguin}\ \emph {et~al.}(2007)\citenamefont
  {Feiguin}, \citenamefont {Trebst}, \citenamefont {Ludwig}, \citenamefont
  {Troyer}, \citenamefont {Kitaev}, \citenamefont {Wang},\ and\ \citenamefont
  {Freedman}}]{Feiguin:2006ydp}%
  \BibitemOpen
  \bibfield  {author} {\bibinfo {author} {\bibfnamefont {A.}~\bibnamefont
  {Feiguin}}, \bibinfo {author} {\bibfnamefont {S.}~\bibnamefont {Trebst}},
  \bibinfo {author} {\bibfnamefont {A.~W.~W.}\ \bibnamefont {Ludwig}}, \bibinfo
  {author} {\bibfnamefont {M.}~\bibnamefont {Troyer}}, \bibinfo {author}
  {\bibfnamefont {A.}~\bibnamefont {Kitaev}}, \bibinfo {author} {\bibfnamefont
  {Z.}~\bibnamefont {Wang}}, \ and\ \bibinfo {author} {\bibfnamefont {M.~H.}\
  \bibnamefont {Freedman}},\ }\href {\doibase 10.1103/PhysRevLett.98.160409}
  {\bibfield  {journal} {\bibinfo  {journal} {Phys. Rev. Lett.}\ }\textbf
  {\bibinfo {volume} {98}},\ \bibinfo {pages} {160409} (\bibinfo {year}
  {2007})},\ \Eprint {http://arxiv.org/abs/cond-mat/0612341}
  {arXiv:cond-mat/0612341} \BibitemShut {NoStop}%
\bibitem [{\citenamefont {Aasen}\ \emph {et~al.}(2016)\citenamefont {Aasen},
  \citenamefont {Mong},\ and\ \citenamefont {Fendley}}]{Aasen:2016dop}%
  \BibitemOpen
  \bibfield  {author} {\bibinfo {author} {\bibfnamefont {D.}~\bibnamefont
  {Aasen}}, \bibinfo {author} {\bibfnamefont {R.~S.~K.}\ \bibnamefont {Mong}},
  \ and\ \bibinfo {author} {\bibfnamefont {P.}~\bibnamefont {Fendley}},\ }\href
  {\doibase 10.1088/1751-8113/49/35/354001} {\bibfield  {journal} {\bibinfo
  {journal} {J. Phys. A}\ }\textbf {\bibinfo {volume} {49}},\ \bibinfo {pages}
  {354001} (\bibinfo {year} {2016})},\ \Eprint
  {http://arxiv.org/abs/1601.07185} {arXiv:1601.07185 [cond-mat.stat-mech]}
  \BibitemShut {NoStop}%
\bibitem [{\citenamefont {Aasen}\ \emph {et~al.}(2020)\citenamefont {Aasen},
  \citenamefont {Fendley},\ and\ \citenamefont {Mong}}]{Aasen:2020jwb}%
  \BibitemOpen
  \bibfield  {author} {\bibinfo {author} {\bibfnamefont {D.}~\bibnamefont
  {Aasen}}, \bibinfo {author} {\bibfnamefont {P.}~\bibnamefont {Fendley}}, \
  and\ \bibinfo {author} {\bibfnamefont {R.~S.~K.}\ \bibnamefont {Mong}},\
  }\href@noop {} {\  (\bibinfo {year} {2020})},\ \Eprint
  {http://arxiv.org/abs/2008.08598} {arXiv:2008.08598 [cond-mat.stat-mech]}
  \BibitemShut {NoStop}%
\bibitem [{\citenamefont {Lootens}\ \emph {et~al.}(2023)\citenamefont
  {Lootens}, \citenamefont {Delcamp}, \citenamefont {Ortiz},\ and\
  \citenamefont {Verstraete}}]{Lootens:2021tet}%
  \BibitemOpen
  \bibfield  {author} {\bibinfo {author} {\bibfnamefont {L.}~\bibnamefont
  {Lootens}}, \bibinfo {author} {\bibfnamefont {C.}~\bibnamefont {Delcamp}},
  \bibinfo {author} {\bibfnamefont {G.}~\bibnamefont {Ortiz}}, \ and\ \bibinfo
  {author} {\bibfnamefont {F.}~\bibnamefont {Verstraete}},\ }\href {\doibase
  10.1103/PRXQuantum.4.020357} {\bibfield  {journal} {\bibinfo  {journal} {PRX
  Quantum}\ }\textbf {\bibinfo {volume} {4}},\ \bibinfo {pages} {020357}
  (\bibinfo {year} {2023})},\ \Eprint {http://arxiv.org/abs/2112.09091}
  {arXiv:2112.09091 [quant-ph]} \BibitemShut {NoStop}%
\bibitem [{\citenamefont {Lootens}\ \emph {et~al.}(2024)\citenamefont
  {Lootens}, \citenamefont {Delcamp},\ and\ \citenamefont
  {Verstraete}}]{Lootens:2022avn}%
  \BibitemOpen
  \bibfield  {author} {\bibinfo {author} {\bibfnamefont {L.}~\bibnamefont
  {Lootens}}, \bibinfo {author} {\bibfnamefont {C.}~\bibnamefont {Delcamp}}, \
  and\ \bibinfo {author} {\bibfnamefont {F.}~\bibnamefont {Verstraete}},\
  }\href {\doibase 10.1103/PRXQuantum.5.010338} {\bibfield  {journal} {\bibinfo
   {journal} {PRX Quantum}\ }\textbf {\bibinfo {volume} {5}},\ \bibinfo {pages}
  {010338} (\bibinfo {year} {2024})},\ \Eprint
  {http://arxiv.org/abs/2211.03777} {arXiv:2211.03777 [quant-ph]} \BibitemShut
  {NoStop}%
\bibitem [{\citenamefont {Inamura}(2022)}]{Inamura:2021szw}%
  \BibitemOpen
  \bibfield  {author} {\bibinfo {author} {\bibfnamefont {K.}~\bibnamefont
  {Inamura}},\ }\href {\doibase 10.1007/JHEP03(2022)036} {\bibfield  {journal}
  {\bibinfo  {journal} {JHEP}\ }\textbf {\bibinfo {volume} {03}},\ \bibinfo
  {pages} {036} (\bibinfo {year} {2022})},\ \Eprint
  {http://arxiv.org/abs/2110.12882} {arXiv:2110.12882 [cond-mat.str-el]}
  \BibitemShut {NoStop}%
\bibitem [{\citenamefont {Delcamp}\ and\ \citenamefont
  {Tiwari}(2024)}]{Delcamp:2023kew}%
  \BibitemOpen
  \bibfield  {author} {\bibinfo {author} {\bibfnamefont {C.}~\bibnamefont
  {Delcamp}}\ and\ \bibinfo {author} {\bibfnamefont {A.}~\bibnamefont
  {Tiwari}},\ }\href {\doibase 10.21468/SciPostPhys.16.4.110} {\bibfield
  {journal} {\bibinfo  {journal} {SciPost Phys.}\ }\textbf {\bibinfo {volume}
  {16}},\ \bibinfo {pages} {110} (\bibinfo {year} {2024})},\ \Eprint
  {http://arxiv.org/abs/2301.01259} {arXiv:2301.01259 [hep-th]} \BibitemShut
  {NoStop}%
\bibitem [{\citenamefont {Inamura}\ and\ \citenamefont
  {Ohmori}(2023)}]{Inamura:2023qzl}%
  \BibitemOpen
  \bibfield  {author} {\bibinfo {author} {\bibfnamefont {K.}~\bibnamefont
  {Inamura}}\ and\ \bibinfo {author} {\bibfnamefont {K.}~\bibnamefont
  {Ohmori}},\ }\href@noop {} {\  (\bibinfo {year} {2023})},\ \Eprint
  {http://arxiv.org/abs/2305.05774} {arXiv:2305.05774 [cond-mat.str-el]}
  \BibitemShut {NoStop}%
\bibitem [{\citenamefont {Chatterjee}\ \emph {et~al.}(2024)\citenamefont
  {Chatterjee}, \citenamefont {Aksoy},\ and\ \citenamefont
  {Wen}}]{Chatterjee:2024ych}%
  \BibitemOpen
  \bibfield  {author} {\bibinfo {author} {\bibfnamefont {A.}~\bibnamefont
  {Chatterjee}}, \bibinfo {author} {\bibfnamefont {O.~M.}\ \bibnamefont
  {Aksoy}}, \ and\ \bibinfo {author} {\bibfnamefont {X.-G.}\ \bibnamefont
  {Wen}},\ }\href@noop {} {\  (\bibinfo {year} {2024})},\ \Eprint
  {http://arxiv.org/abs/2405.05331} {arXiv:2405.05331 [cond-mat.str-el]}
  \BibitemShut {NoStop}%
\bibitem [{\citenamefont {Schafer-Nameki}(2024)}]{Schafer-Nameki:2023jdn}%
  \BibitemOpen
  \bibfield  {author} {\bibinfo {author} {\bibfnamefont {S.}~\bibnamefont
  {Schafer-Nameki}},\ }\href {\doibase 10.1016/j.physrep.2024.01.007}
  {\bibfield  {journal} {\bibinfo  {journal} {Phys. Rept.}\ }\textbf {\bibinfo
  {volume} {1063}},\ \bibinfo {pages} {1} (\bibinfo {year} {2024})},\ \Eprint
  {http://arxiv.org/abs/2305.18296} {arXiv:2305.18296 [hep-th]} \BibitemShut
  {NoStop}%
\bibitem [{\citenamefont {Shao}(2023)}]{Shao:2023gho}%
  \BibitemOpen
  \bibfield  {author} {\bibinfo {author} {\bibfnamefont {S.-H.}\ \bibnamefont
  {Shao}},\ }\href@noop {} {\  (\bibinfo {year} {2023})},\ \Eprint
  {http://arxiv.org/abs/2308.00747} {arXiv:2308.00747 [hep-th]} \BibitemShut
  {NoStop}%
\bibitem [{\citenamefont {Vanhove}\ \emph {et~al.}(2018)\citenamefont
  {Vanhove}, \citenamefont {Bal}, \citenamefont {Williamson}, \citenamefont
  {Bultinck}, \citenamefont {Haegeman},\ and\ \citenamefont
  {Verstraete}}]{Vanhove:2018wlb}%
  \BibitemOpen
  \bibfield  {author} {\bibinfo {author} {\bibfnamefont {R.}~\bibnamefont
  {Vanhove}}, \bibinfo {author} {\bibfnamefont {M.}~\bibnamefont {Bal}},
  \bibinfo {author} {\bibfnamefont {D.~J.}\ \bibnamefont {Williamson}},
  \bibinfo {author} {\bibfnamefont {N.}~\bibnamefont {Bultinck}}, \bibinfo
  {author} {\bibfnamefont {J.}~\bibnamefont {Haegeman}}, \ and\ \bibinfo
  {author} {\bibfnamefont {F.}~\bibnamefont {Verstraete}},\ }\href {\doibase
  10.1103/PhysRevLett.121.177203} {\bibfield  {journal} {\bibinfo  {journal}
  {Phys. Rev. Lett.}\ }\textbf {\bibinfo {volume} {121}},\ \bibinfo {pages}
  {177203} (\bibinfo {year} {2018})},\ \Eprint
  {http://arxiv.org/abs/1801.05959} {arXiv:1801.05959 [quant-ph]} \BibitemShut
  {NoStop}%
\bibitem [{\citenamefont {Chatterjee}\ \emph {et~al.}(2022)\citenamefont
  {Chatterjee}, \citenamefont {Ji},\ and\ \citenamefont
  {Wen}}]{Chatterjee:2022jll}%
  \BibitemOpen
  \bibfield  {author} {\bibinfo {author} {\bibfnamefont {A.}~\bibnamefont
  {Chatterjee}}, \bibinfo {author} {\bibfnamefont {W.}~\bibnamefont {Ji}}, \
  and\ \bibinfo {author} {\bibfnamefont {X.-G.}\ \bibnamefont {Wen}},\
  }\href@noop {} {\  (\bibinfo {year} {2022})},\ \Eprint
  {http://arxiv.org/abs/2212.14432} {arXiv:2212.14432 [cond-mat.str-el]}
  \BibitemShut {NoStop}%
\bibitem [{\citenamefont {Chatterjee}\ and\ \citenamefont
  {Wen}(2023)}]{Chatterjee:2022tyg}%
  \BibitemOpen
  \bibfield  {author} {\bibinfo {author} {\bibfnamefont {A.}~\bibnamefont
  {Chatterjee}}\ and\ \bibinfo {author} {\bibfnamefont {X.-G.}\ \bibnamefont
  {Wen}},\ }\href {\doibase 10.1103/PhysRevB.108.075105} {\bibfield  {journal}
  {\bibinfo  {journal} {Phys. Rev. B}\ }\textbf {\bibinfo {volume} {108}},\
  \bibinfo {pages} {075105} (\bibinfo {year} {2023})},\ \Eprint
  {http://arxiv.org/abs/2205.06244} {arXiv:2205.06244 [cond-mat.str-el]}
  \BibitemShut {NoStop}%
\bibitem [{\citenamefont {Wen}\ and\ \citenamefont
  {Potter}(2023{\natexlab{b}})}]{Wen:2023otf}%
  \BibitemOpen
  \bibfield  {author} {\bibinfo {author} {\bibfnamefont {R.}~\bibnamefont
  {Wen}}\ and\ \bibinfo {author} {\bibfnamefont {A.~C.}\ \bibnamefont
  {Potter}},\ }\href@noop {} {\  (\bibinfo {year} {2023}{\natexlab{b}})},\
  \Eprint {http://arxiv.org/abs/2311.00050} {arXiv:2311.00050
  [cond-mat.str-el]} \BibitemShut {NoStop}%
\bibitem [{\citenamefont {Huang}\ and\ \citenamefont
  {Cheng}(2023)}]{Huang:2023pyk}%
  \BibitemOpen
  \bibfield  {author} {\bibinfo {author} {\bibfnamefont {S.-J.}\ \bibnamefont
  {Huang}}\ and\ \bibinfo {author} {\bibfnamefont {M.}~\bibnamefont {Cheng}},\
  }\href@noop {} {\  (\bibinfo {year} {2023})},\ \Eprint
  {http://arxiv.org/abs/2310.16878} {arXiv:2310.16878 [cond-mat.str-el]}
  \BibitemShut {NoStop}%
\bibitem [{\citenamefont {Moradi}\ \emph {et~al.}(2023)\citenamefont {Moradi},
  \citenamefont {Moosavian},\ and\ \citenamefont {Tiwari}}]{Moradi:2022lqp}%
  \BibitemOpen
  \bibfield  {author} {\bibinfo {author} {\bibfnamefont {H.}~\bibnamefont
  {Moradi}}, \bibinfo {author} {\bibfnamefont {S.~F.}\ \bibnamefont
  {Moosavian}}, \ and\ \bibinfo {author} {\bibfnamefont {A.}~\bibnamefont
  {Tiwari}},\ }\href {\doibase 10.21468/SciPostPhysCore.6.4.066} {\bibfield
  {journal} {\bibinfo  {journal} {SciPost Phys. Core}\ }\textbf {\bibinfo
  {volume} {6}},\ \bibinfo {pages} {066} (\bibinfo {year} {2023})},\ \Eprint
  {http://arxiv.org/abs/2207.10712} {arXiv:2207.10712 [cond-mat.str-el]}
  \BibitemShut {NoStop}%
\bibitem [{\citenamefont {Chen}\ \emph {et~al.}(2022)\citenamefont {Chen},
  \citenamefont {Zhang}, \citenamefont {Ji}, \citenamefont {Shen},
  \citenamefont {Wang}, \citenamefont {Zeng},\ and\ \citenamefont
  {Hung}}]{Chen:2022wvy}%
  \BibitemOpen
  \bibfield  {author} {\bibinfo {author} {\bibfnamefont {L.}~\bibnamefont
  {Chen}}, \bibinfo {author} {\bibfnamefont {H.}~\bibnamefont {Zhang}},
  \bibinfo {author} {\bibfnamefont {K.}~\bibnamefont {Ji}}, \bibinfo {author}
  {\bibfnamefont {C.}~\bibnamefont {Shen}}, \bibinfo {author} {\bibfnamefont
  {R.}~\bibnamefont {Wang}}, \bibinfo {author} {\bibfnamefont {X.}~\bibnamefont
  {Zeng}}, \ and\ \bibinfo {author} {\bibfnamefont {L.-Y.}\ \bibnamefont
  {Hung}},\ }\href@noop {} {\  (\bibinfo {year} {2022})},\ \Eprint
  {http://arxiv.org/abs/2210.12127} {arXiv:2210.12127 [hep-th]} \BibitemShut
  {NoStop}%
\bibitem [{\citenamefont {Fechisin}\ \emph {et~al.}(2023)\citenamefont
  {Fechisin}, \citenamefont {Tantivasadakarn},\ and\ \citenamefont
  {Albert}}]{Fechisin:2023dkj}%
  \BibitemOpen
  \bibfield  {author} {\bibinfo {author} {\bibfnamefont {C.}~\bibnamefont
  {Fechisin}}, \bibinfo {author} {\bibfnamefont {N.}~\bibnamefont
  {Tantivasadakarn}}, \ and\ \bibinfo {author} {\bibfnamefont {V.~V.}\
  \bibnamefont {Albert}},\ }\href@noop {} {\  (\bibinfo {year} {2023})},\
  \Eprint {http://arxiv.org/abs/2312.09272} {arXiv:2312.09272
  [cond-mat.str-el]} \BibitemShut {NoStop}%
\bibitem [{\citenamefont {Seifnashri}\ and\ \citenamefont
  {Shao}(2024)}]{Seifnashri:2024dsd}%
  \BibitemOpen
  \bibfield  {author} {\bibinfo {author} {\bibfnamefont {S.}~\bibnamefont
  {Seifnashri}}\ and\ \bibinfo {author} {\bibfnamefont {S.-H.}\ \bibnamefont
  {Shao}},\ }\href@noop {} {\  (\bibinfo {year} {2024})},\ \Eprint
  {http://arxiv.org/abs/2404.01369} {arXiv:2404.01369 [cond-mat.str-el]}
  \BibitemShut {NoStop}%
\end{thebibliography}%


\end{document}